\documentclass[conference,compsoc]{IEEEtran}
\IEEEoverridecommandlockouts

\usepackage{tikz}

\usepackage{amsmath,amssymb,amsfonts}
\usepackage{algorithmic}
\usepackage{graphicx}
\usepackage{textcomp}
\usepackage{xcolor}
\usepackage{xspace}
\usepackage{hyperref}
\usepackage{listings}
\newcommand{\codeinline}[1]{\lstinline[language=python,basicstyle=\ttfamily,breaklines=true,breakatwhitespace=true,showstringspaces=false]{#1}}
\usepackage{enumitem}
\usepackage{multirow}
\usepackage{booktabs}
\usepackage{array}
\usepackage{subcaption}
\usepackage{pifont}
\usepackage{fancyvrb}
\usepackage{optidef}
\usepackage{makecell}
\usepackage{listings}

\usepackage{multicol}
\usepackage{adjustbox}
\usepackage{colortbl}
\usepackage{bm}

\usepackage[acronym]{glossaries}

\hypersetup{%
  colorlinks=true,
  linkcolor=black,
  citecolor=black,
  urlcolor=black
}

\newacronym{ML}{ML}{Machine Learning}
\newacronym{DL}{DL}{Deep Learning}
\newacronym{CFG}{CFG}{Control Flow Graph}
\newacronym{FCG}{FCG}{Function Call Graph}
\newacronym{PDG}{PDG}{Program Dependency Graph}
\newacronym{LLM}{LLM}{Large Language Model}
\newacronym{IOC}{IoC}{Indicators of Compromise}

\graphicspath{{figures/}}

\newcolumntype{C}[1]{>{\centering\let\newline\\\arraybackslash\hspace{0pt}}m{#1}}

\makeatletter
\AtBeginDocument{\let\c@listing\c@lstlisting}
\makeatother

\def\BibTeX{{\rm B\kern-.05em{\sc i\kern-.025em b}\kern-.08em
    T\kern-.1667em\lower.7ex\hbox{E}\kern-.125emX}}

\makeatletter
\newcommand{\linebreakand}{%
    \end{@IEEEauthorhalign}
    \hfill\mbox{}\par
    \mbox{}\hfill\begin{@IEEEauthorhalign}
}
\makeatother

\begin{document}

\title{One Detector Fits All: \\ Robust and Adaptive Detection of Malicious Packages from PyPI to Enterprises}

\author{
\IEEEauthorblockN{Biagio Montaruli}
\IEEEauthorblockA{\textit{SAP} \& \textit{EURECOM}\\
Nice, France \\
biagio.montaruli@eurecom.fr}
\and
\IEEEauthorblockN{Luca Compagna}
\IEEEauthorblockA{\textit{Endor Labs} \\
Palo Alto, USA \\
lcompagna@endor.ai}
\and
\IEEEauthorblockN{Serena Elisa Ponta}
\IEEEauthorblockA{\textit{SAP} \\
Mougins, France \\
serena.ponta@sap.com}
\and
\IEEEauthorblockN{Davide Balzarotti}
\IEEEauthorblockA{\textit{EURECOM} \\
Biot, France \\
davide.balzarotti@eurecom.fr}
}

\maketitle

\newcommand{\sota}{state-of-the-art\xspace}
\newcommand{\diff}[2]{\frac{\partial #1}{\partial #2}}
\newcommand{\vct}[1]{\ensuremath{\boldsymbol{#1}}}
\newcommand{\mat}[1]{\ensuremath{\mathbf{#1}}}
\newcommand{\set}[1]{\ensuremath{\mathcal{#1}}}
\newcommand{\con}[1]{\ensuremath{\mathsf{#1}}}
\newcommand{\tsum}{\ensuremath{\textstyle \sum}}
\newcommand{\T}{\ensuremath{\top}}
\newcommand{\ind}[1]{\ensuremath{\mathbbm 1_{#1}}}
\newcommand{\argmax}{\operatornamewithlimits{\arg\,\max}}
\newcommand{\erf}{\text{erf}}
\newcommand{\sign}{\text{sign}}
\newcommand{\argmin}{\operatornamewithlimits{\arg\,\min}}
\newcommand{\bmat}[1]{\begin{bmatrix}#1\end{bmatrix}}
\newcommand{\myparagraph}[1]{\noindent \textbf{#1}}
\newcommand{\myparagraphlb}[1]{\noindent \newline \textbf{#1}}
\newcommand{\mysubparagraph}[1]{\noindent \underline{\textit{#1}}}
\newcommand{\ie}{i.e.\xspace}
\newcommand{\eg}{e.g.\xspace}
\newcommand{\etc}{etc.\xspace}
\newcommand{\aka}{a.k.a.\xspace}
\newcommand{\wrt}{w.r.t.\xspace}
\newcommand{\etal}{\emph{et al.}\xspace}

\newcommand{\ioc}{Indicators Of Compromise\xspace}
\newcommand{\cc}{Command and Control\xspace}

\newcommand{\takeaways}[1]{
\noindent \fcolorbox{gray}{white}{\begin{minipage}[c]{0.96\columnwidth}
\textbf{Takeaways}: #1
\end{minipage}}
}
\newcommand{\takeawaysNoTitle}[1]{
\noindent \fcolorbox{gray}{white}{\begin{minipage}[c]{0.96\columnwidth}
#1
\end{minipage}}
}

\begin{abstract}
The rise of supply chain attacks via malicious Python packages demands robust detection solutions.
Current approaches, however, overlook two critical challenges: robustness against adversarial source code transformations and adaptability to the varying false positive rate (FPR) requirements of different actors,
from repository maintainers (requiring low FPR) to enterprise security teams (higher FPR tolerance).

We introduce a robust detector capable of seamless integration into both public repositories like PyPI and enterprise ecosystems.
To ensure robustness, we propose a novel methodology for generating adversarial packages using fine-grained code obfuscation.
Combining these with adversarial training (AT) enhances detector robustness by 2.5x.
We comprehensively evaluate AT effectiveness by testing our detector against 122,398 packages collected daily from PyPI over 80 days, showing that AT needs careful application: it makes the detector more robust to obfuscations and allows finding 10\% more obfuscated packages, but slightly decreases performance on non-obfuscated packages.

We demonstrate production adaptability of our detector via two case studies: (i) one for PyPI maintainers (tuned at 0.1\% FPR) and (ii) one for enterprise teams (tuned at 10\% FPR).
In the former, we analyze 91,949 packages collected from PyPI over 37 days, achieving a daily detection rate of 2.48 malicious packages with only 2.18 false positives.
In the latter, we analyze 1,596 packages adopted by a multinational software company, obtaining only 1.24 false positives daily.
These results show that our detector can be seamlessly integrated into both public repositories like PyPI and enterprise ecosystems, ensuring a very low time budget of a few minutes to review the false positives.

Overall, we uncovered 346 malicious packages, now reported to the community.
\end{abstract}


\section{Introduction}

Supply chain attacks are a growing threat to the software industry. 
According to the \emph{2024 State of the Software Supply Chain} report published by Sonatype,
the number of malicious packages discovered in the wild had a yearly increase of 156\%, with 512,847 new malicious packages identified in 2024~\cite{sonatype2024}.
Recently, PyPI, one of the main ecosystems for Python packages, has faced a growing number of attacks that even led to a temporary halt in the creation of new projects and the registration of new users~\cite{Checkmarx2024BlogPost,Goodin2024ArsTechnicaArticle}.

To prevent the spread of malicious software packages before they cause considerable harm, it is crucial to accurately and swiftly analyze newly uploaded packages.
Yet, the detection of malicious packages is still an open problem in real-world scenarios~\cite{sonatype2024}.
While this has been the subject of several recent studies~\cite{Ladisa2023ACSAC,zhang2023cerebro,malwukong,oscar_ase24,Halder2024MeMPtec,huang2024donapi}, the current state-of-the-art overlooks two important open challenges.
First, existing approaches do not consider robustness to adversarial transformations, which is a crucial factor in an adversarial environment.
Second, researchers have not yet studied how a given solution can be adapted to different stakeholders, who in turn have different requirements in terms of acceptable false positive rates (FPRs) and tolerance for false negatives (FNs).
For instance, repository maintainers with scarce resources might prioritize low FPR, while dedicated enterprise security teams might tolerate higher FPR in exchange for better security guarantees.

\begin{figure*}[!t]
    \centering
    \includegraphics[width=0.88\textwidth]{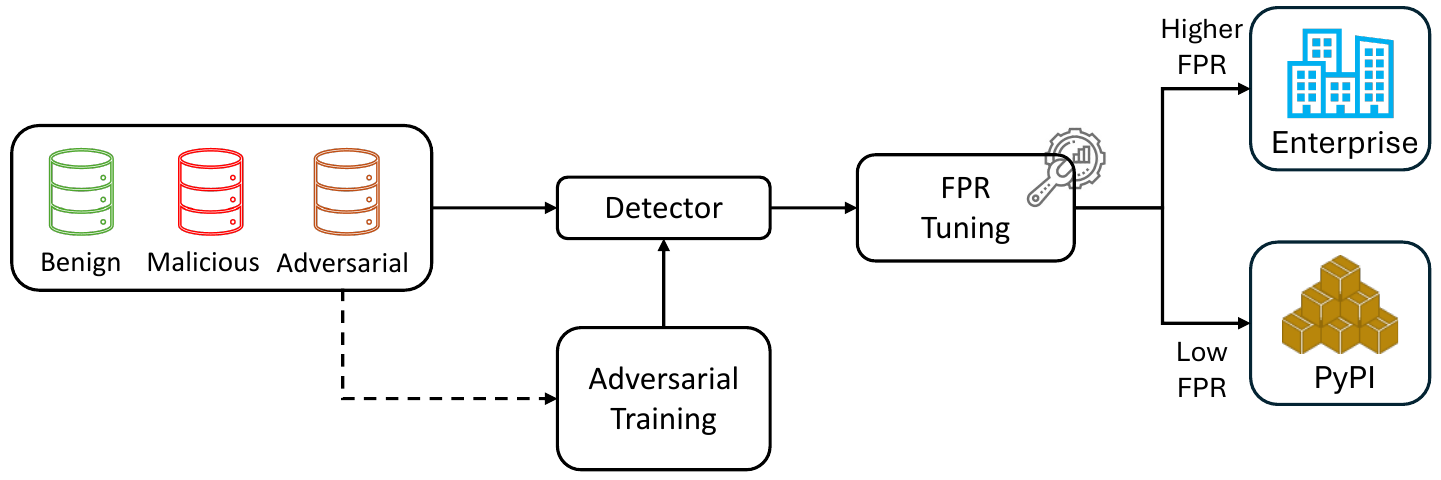}
    \caption{
    Proposed approach: we design a robust and adaptive detector of malicious Python packages, which can be tuned to the needs of different actors in the software supply chain,
    from PyPI maintainers to enterprise security teams.
    }
    \label{fig:approach}
\end{figure*}

\vspace{0.2cm}
\noindent \textbf{Adversarial Setting} -- Several detectors based on static signatures and machine learning techniques 
have been proposed to identify malicious packages in popular ecosystems such as PyPI and NPM~\cite{Ladisa2023ACSAC,zhang2023cerebro,malwukong,oscar_ase24,Halder2024MeMPtec,huang2024donapi}.
However, none of the previous studies have systematically evaluated the robustness of the current malicious package detectors against adversarial attacks, 
\ie, carefully crafted inputs that are designed to mislead the system into making incorrect predictions~\cite{biggio2018wild,Demontis2019YesML}.
This is a significant gap in the literature, since a clear understanding of adversarial transformations would open the door to more effective countermeasures, for instance, by leveraging adversarial training (AT)~\cite{biggio2018wild},
a well-known technique that augments training data with adversarial examples, thereby enabling the detector to withstand the corresponding evasive attack patterns at test time.
However, to the best of our knowledge, no prior study has thoroughly evaluated the effectiveness of AT in the context of malicious package detection.

\vspace{0.2cm}
\noindent \textbf{Operational Tuning} -- 
Existing solutions were not designed to address the needs of different actors in the software supply chain, who have different requirements in terms of detection and false positive rates.
For instance, repository maintainers require a very low FPR due to the high volume of packages uploaded daily, but can tolerate false negatives~\cite{Vu2023BadSnakes}.
It is worth mentioning that in 2020, PyPI introduced a malware scanning pilot, but it was discontinued two years later due to the overwhelming number of false positives~\cite{Vu2023BadSnakes}.
This clearly indicates that current solutions fail to meet the needs of repository maintainers.
On the other hand, security engineers in software enterprises need to monitor only a small subset of the available packages (typically those actively used within their organization),
and therefore can tolerate a higher FPR in exchange for better detection performance.
Hence, a flexible solution that can be easily customized to the needs of different actors in the software supply chain is highly desirable.

\vspace{0.2cm} 
\noindent \textbf{Contributions} -- 
In this work, we aim to fill the above-mentioned gaps by introducing a novel robust approach (see \autoref{fig:approach}) that can be easily customized to the needs of different actors in the software supply chain by tuning the classification threshold to the desired FPR.
We extensively evaluate our solution on real-world datasets to demonstrate its robustness and adaptability in real-world scenarios.
To this end, our study makes several contributions.

First, we propose a novel methodology to generate \emph{adversarial packages} -- malicious packages generated by leveraging functionality-preserving adversarial transformations of the source code,
\ie, transformations that modify the source code of malicious packages without compromising their malicious intent, 
while preserving syntactic and semantic correctness~\cite{ArpUsenix2022,demetrio2021tifs}.
Specifically, we focus on the PyPI ecosystem and propose a novel set of transformations that can be applied to Python source code.
We show that our transformations effectively bypass several state-of-the-art detectors with an 87.42\% success rate.

Second, we comprehensively evaluate the effectiveness of adversarial training (AT) in the context of malicious package detection
by experimenting on a real-world dataset of 122,398 packages collected daily from PyPI over a period of 80 days.
Our results show that AT has a double-edged effect.
On the one hand, it significantly improves the robustness of our detector by 2.5$\times$ against the adversarial transformations and allows finding 6 (+10\%) more obfuscated packages compared to the baseline detector.
On the other hand, it negatively impacts, even if slightly, the performance on non-obfuscated packages (+2 malicious packages detected by the baseline w.r.t. the AT-based detector).
To the best of our knowledge, this is the first study that proposes a systematic methodology to evaluate the robustness 
of malicious package detectors against adversarial attacks in the problem space~\cite{ArpUsenix2022}, and a thorough evaluation of AT in this domain.

Third, we thoroughly evaluate the adaptability of our solution on two real-world case studies:
a first study addressing the needs of PyPI maintainers (tuned at 0.1\% FPR) and a second one conducted in collaboration with an enterprise security team (tuned at 10\% FPR).
Both case studies lasted for 37 days.
In the first study, we analyzed 91,949 packages collected from PyPI and our solution was able to detect an average of 2.48 malicious packages per day,
with only 2 false positives to analyze per day, thus ensuring a very low effort for the PyPI maintainers to vet the results.
In the second case, our solution was used to monitor 1,596 packages adopted by a large multinational software company, and, even if using a very high FPR of 10\%,
we achieved an average of 1.24 false positives to analyze per day, which implies only a few minutes of work for an enterprise security team.

Overall, we identified and reported to the community \textbf{346} malicious Python packages.

Finally, to foster further research on this topic we release our code and data to the research community at this link:
\textbf{\url{https://github.com/SAP-samples/robust-pypi-detector}}.

\section{Methodology}\label{sec:adv_pkg}

This section describes the proposed methodology for generating adversarial packages.
These packages serve two distinct purposes:
(i) to assess the adversarial robustness of the detectors evaluated in our experiments, and 
(ii) to improve their robustness through adversarial training (AT) by incorporating adversarial packages into the training process.
To this end, we first describe the adversarial transformations 
used to generate the adversarial packages (Section~\ref{sec:transformations}), 
and then detail the process of adversarial training (Section~\ref{sec:adversarial_training}).

\subsection{Adversarial Transformations of Source Code}\label{sec:transformations}
The adversarial transformations adopted in our work are summarized in \autoref{table:transformations}.
The list consists of a set of fine-grained and functionality-preserving operations based on common code obfuscation techniques proposed by Schrittwieser et al.~\cite{Schrittwieser2017ObfuscationSurvey}. 
Their goal is to modify the source code of a given package without compromising its (malicious) functionality, preserving the syntactic and semantic correctness of the code while making it more challenging for static analyzers and detectors based on static features to identify the malicious content.

In particular, in our work we focus on obfuscation of security-relevant strings (\ie, IPs, URLs, system commands), API calls, and techniques that modify the structure of the source code to evade detection.
To this end, we leveraged the categorization proposed by Ladisa \etal~\cite{Ladisa2023SCORED}, which classifies the most common obfuscation techniques used by malicious packages into three main categories: data obfuscation, static code transformations, and dynamic code transformations.
Given our focus on static classifiers, we cover the transformations in the first two categories and tailor their implementation to the Python programming language.
To generate adversarial packages, these transformations are combined together and optimized against the target detector by leveraging a state-of-the-art black-box optimization algorithm (described in Section~\ref{sec:adversarial_training}).

Moreover, we remark that, while our transformations are built on insights from previous research~\cite{Ladisa2023SCORED,Schrittwieser2017ObfuscationSurvey}, no prior work has comprehensively evaluated their effectiveness for assessing the adversarial robustness of malicious package detectors.
As for the novelty of our methodology, in this work we introduce a new transformation named \emph{API obfuscation} --  to the best of our knowledge never studied in previous work -- that leverages Python's polymorphic syntax to rewrite API calls in diverse ways to evade detection.
Furthermore, we would like to highlight that all the transformations are functionality-preserving by design, as they leverage built-in functionalities of the Python programming language that preserve the original semantics of the code and maintain syntactic correctness.
Finally, we remark that we have extensively validated the correctness and functionality-preserving nature of our transformations through several unit tests and verified their correctness on some real-world malicious packages.

\begin{table*}[!t]
    \centering
    \begin{adjustbox}{max width=\textwidth}
    \begin{tabular}{c c c}
    \toprule
    \textbf{Category} & \textbf{Transformation} & \textbf{Description} \\
    \toprule
    \multirow{5}{*}{\shortstack{Data \\ Obfuscation}}
    & Encoding & \makecell{Encode the string using a specific encoding scheme, \\ such as Base64, Base32, Base16, or hexadecimal.} \\
    \cline{2-3}
    & Binary Arrays & \makecell{Represent the string as a binary array \\ and manipulate it using bitwise operations.} \\
    \cline{2-3}
    & Data Reordering & \makecell{Split the string into multiple substrings \\ and reorder them at runtime.} \\
    \midrule
    \multirow{5}{*}{\shortstack{Static Code \\ Transformation}}
    & Renaming Identifiers & \makecell{Rename identifiers (e.g., variables and function names) \\ to evade detection by static analysis tools.} \\
    \cline{2-3}
    & \makecell{Useless Code \\ Injection} & \makecell{Inject dead or useless code snippets to hide \\ the malicious content and increase analysis complexity.} \\
    \cline{2-3}
    & API Obfuscation & \makecell{Replace an API import, call, or reference in the source code \\ with a semantically equivalent syntax to evade detection.} \\
    \bottomrule
    \end{tabular}
    \end{adjustbox}
    \vspace{+0.1cm}
    \caption{Summary of the adversarial transformations adopted in this work.
    }
    \label{table:transformations}
\end{table*}

\subsubsection{Data Obfuscation}
This category comprises a set of transformations that modify the way strings are represented within the source code, to obfuscate them from static analysis techniques.
Malicious packages often include hard-coded strings such as URLs or IP addresses that point to a \cc (C\&C) server, or a shell command to execute a reverse shell~\cite{Guo2023EmpiricalStudyPyPI,Ladisa2023SCORED}.
Since these indicators can reveal information about the attacker's techniques and intended goals,
from an attacker's point of view it is crucial to leverage data obfuscation techniques to evade detection and hide details that could expose their identity.
For these reasons, even though these transformations can be applied to any string in the source code, our work focuses specifically on strings representing URLs, IPs and system commands.
Nevertheless, we remark that these transformations can be applied to any string in the source code.

\begin{figure}[t]
    \centering
    \includegraphics[width=\columnwidth]{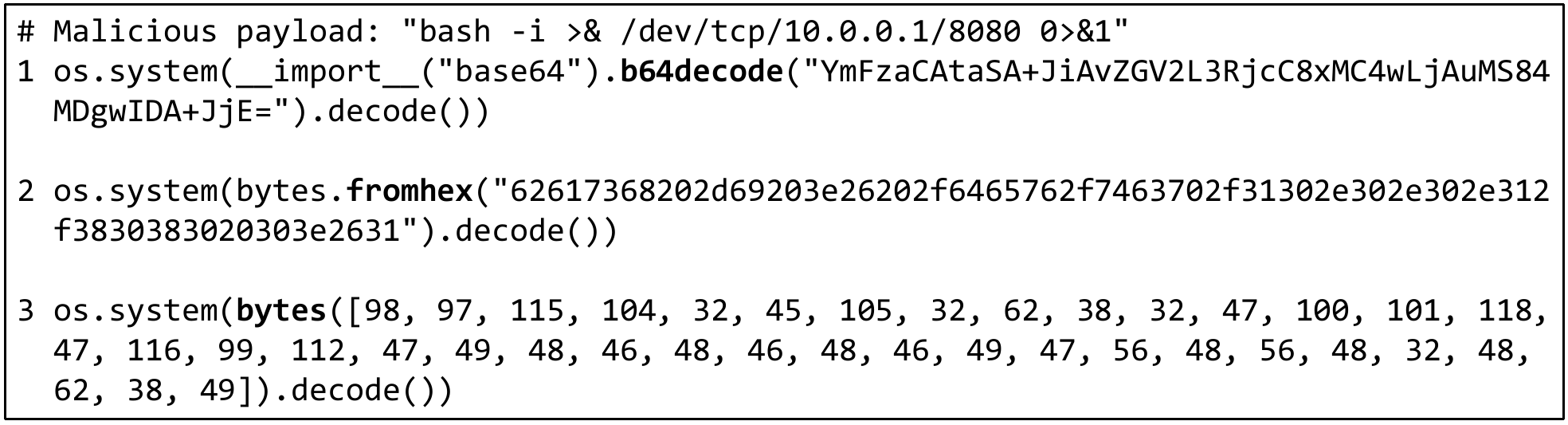}
    \caption{Example of how to obfuscate the malicious payload \codeinline{"bash -i >& /dev/tcp/10.0.0.1/8080 0>&1"} (reverse shell) with the corresponding string encoded in Base64 (line \#1), hexadecimal (line \#2) and byte array representation (line \#3), 
    which is then decoded at runtime and executed using the \texttt{os.system()} function.}
    \label{fig:encoding}
\end{figure}

\myparagraph{Encoding.}
This transformation consists of encoding a given string using a specific encoding scheme, replacing the original string with the encoded version, and decoding it at runtime to retrieve the original content.
Among the main encoding schemes, we consider Base64, Base32, Base16, and hexadecimal encoding, as they have been observed in many real-world attacks~\cite{phylum2023Blog,rlblog2023} and are natively supported by Python.
In our implementation, the encoding scheme is randomly selected for each string to be obfuscated, and the decoding function is imported (if needed) and executed inline.
For instance, \autoref{fig:encoding} shows how the malicious payload \codeinline{"bash -i >& /dev/tcp/10.0.0.1/8080 0>&1"} (a reverse shell), which can be executed using the \verb|os.system()| function,
is replaced with the corresponding encoded string in Base64 (line \#1) or hexadecimal (line \#2), and then decoded at runtime using the related decoding functions (\ie, \codeinline{b64decode()} and \codeinline{fromhex()}).

\myparagraph{Binary Arrays.}
This transformation consists of representing strings as binary arrays.
In this way, attackers can manipulate them using bitwise operations, XOR operations, or with custom encoding schemes to further obfuscate the strings~\cite{Ladisa2023SCORED,Savio2016CodeObfuscation}.
To this end, Python provides the \verb|bytearray()| and \verb|bytes()| functions that can be used to represent and manipulate binary data.
\autoref{fig:encoding} shows how the malicious payload \codeinline{"bash -i >& /dev/tcp/10.0.0.1/8080 0>&1"} is replaced with the corresponding byte array representation (line \#3), which is then decoded and executed at runtime.

\begin{figure}[t]
    \centering
    \includegraphics[width=0.9\columnwidth]{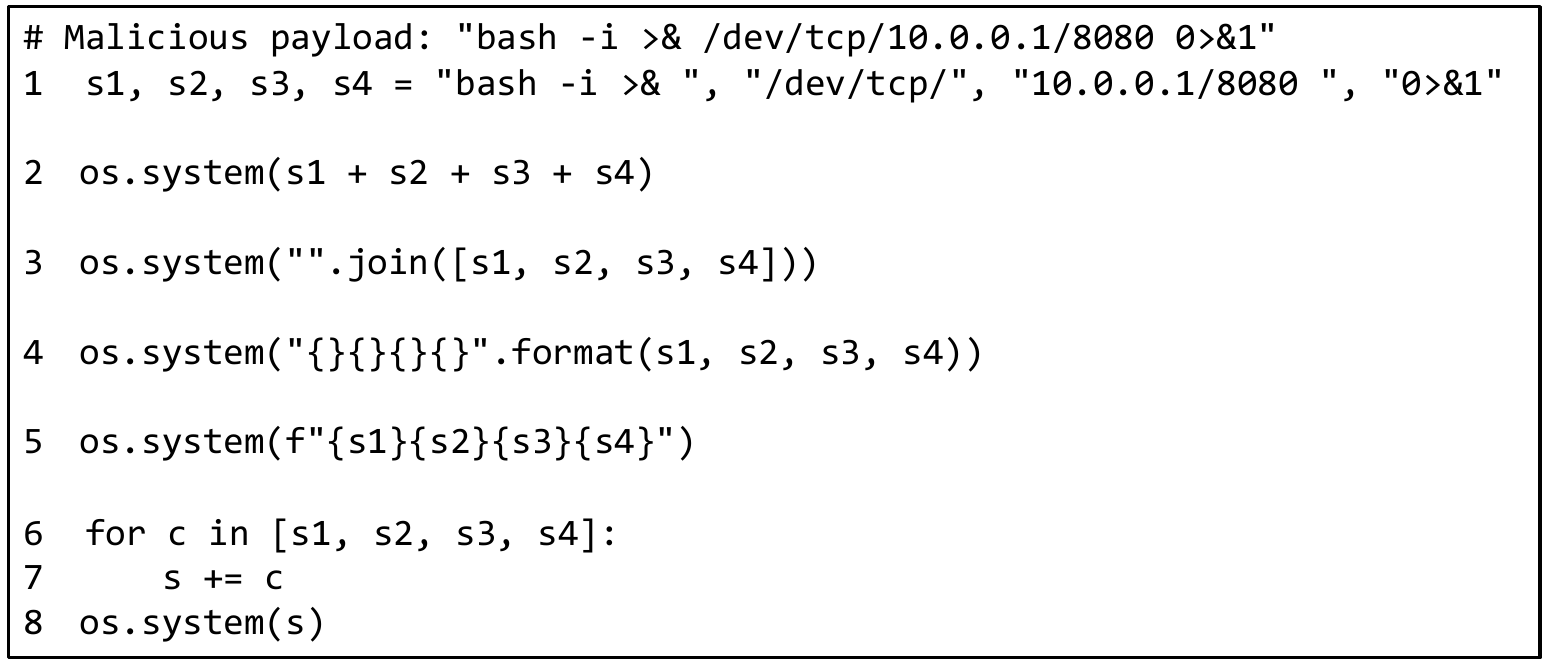}
    \caption{
    Example of how to split the malicious payload \codeinline{"bash -i >& /dev/tcp/10.0.0.1/8080 0>&1"} into multiple substrings 
    and reorder them in several equivalent ways in Python.
    }
    \label{fig:reordering}
\end{figure}

\myparagraph{Data Reordering.}
This transformation involves splitting strings into multiple substrings and reordering them to obfuscate the original content.
Detection is complicated by the fact that programming languages like Python and JavaScript offer many ways to reorder substrings.
For instance, as shown in \autoref{fig:reordering}, the malicious payload \codeinline{"bash -i >& /dev/tcp/10.0.0.1/8080 0>&1"} can be split into multiple substrings (line \#1) and reordered in several equivalent approaches:
by joining the substrings with the \verb|+| operator (line \#2), by using the \verb|join()| function of the \verb|str| class (line \#3), by creating a formatted string using the \verb|format()| function (line \#4), by using the \verb|f-string| syntax (line \#5),
or by concatenating the substrings using a \verb|for| loop (lines \#6–8). 

\subsubsection{Static Code Transformation}
This category includes transformations that obfuscate the source code by modifying its structure without changing its functionality.

\myparagraph{Renaming Identifiers.}
This transformation, observed in real-world attacks~\cite{Unit422023MaliciousPackagesWin}, involves renaming identifiers (\eg, variable, function and class names) in the source code to evade detection by static analysis tools.
For instance, when importing a module or a method such as \texttt{from os import system}, the attacker can rename the \codeinline{system} method to evade detection, as in \texttt{from os import system as \_ssystem}.
In this work, we support renaming identifiers representing security-sensitive modules and related methods such as \codeinline{os.system()}, widely used to execute system commands.

\myparagraph{Useless Code Injection.}
This transformation involves injecting useless code snippets into the source code to conceal the malicious content and make the malicious package appear more similar to benign ones, thereby increasing the complexity of the analysis process.
We implement this transformation by adding comments, whitespace characters, or code snippets that do not affect the package's functionality.

\begin{figure*}[!t]
    \centering
    \includegraphics[width=0.78\textwidth]{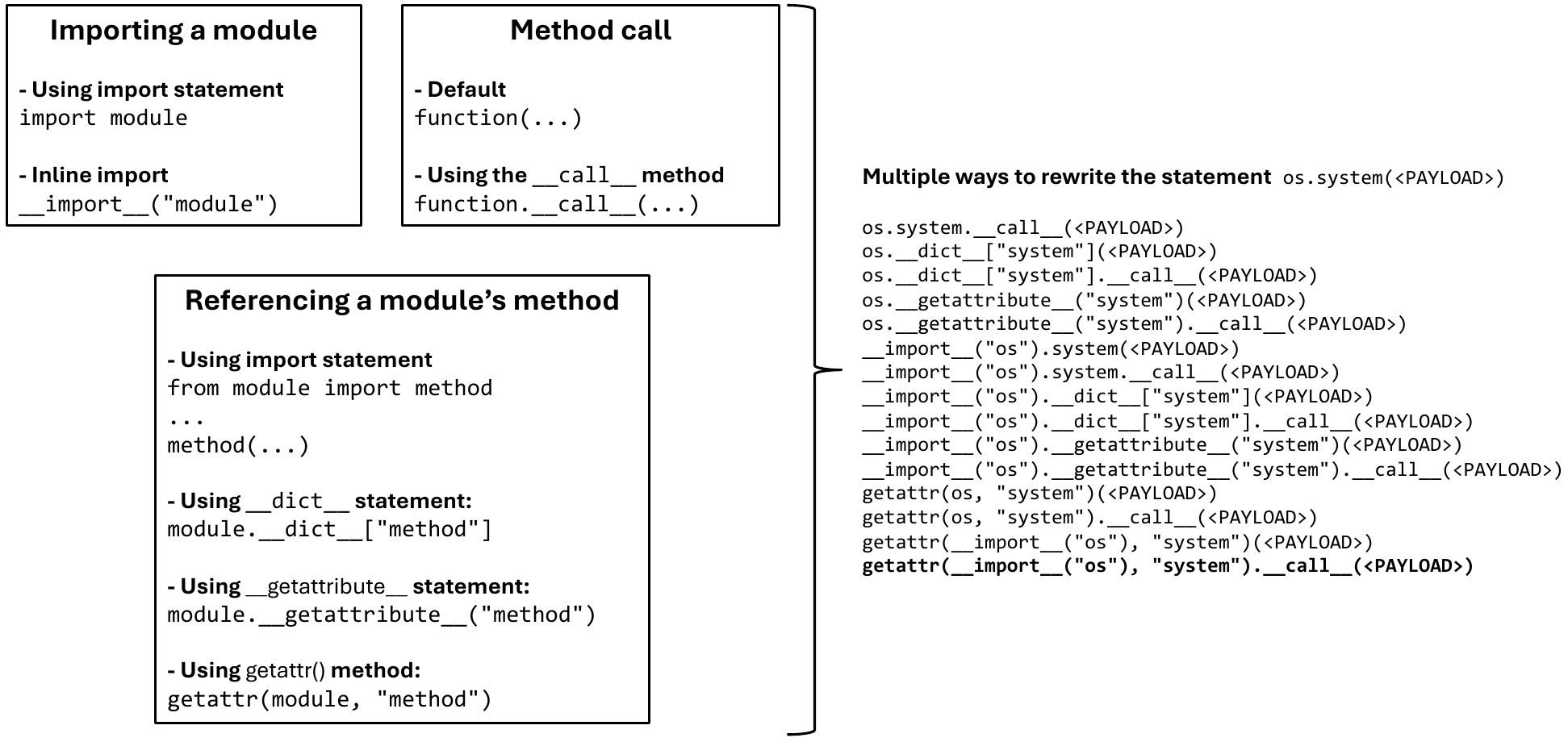}
    \caption{Example of API obfuscation transformation to rewrite a module import, a method call, and a method reference using an alternative but semantically equivalent syntax.}
    \label{fig:api_obfuscation}
\end{figure*}

\myparagraph{API obfuscation.} 
This novel transformation consists of obfuscating API calls in the source code by rewriting them using an alternative but semantically equivalent syntax to avoid detection by static analysis tools, particularly those relying on pattern matching of API calls, such as GuardDog~\cite{datadog2022guarddog}.
Specifically, as shown in \autoref{fig:api_obfuscation}, this transformation leverages the polymorphic nature of Python's syntax to import a module, call a method (or a function), and reference a given method included in a module.
As for modules' imports, the common \texttt{import} statement can be replaced with the \texttt{\_\_import\_\_()} function that allows to import a module dynamically.
For example, the statement \texttt{import os} can be replaced with \texttt{\_\_import\_\_("os")}.
Similarly, the standard way of calling a method, \ie, \codeinline{method(...)}, can be replaced with the equivalent \texttt{method.\_\_call\_\_(...)}.
Finally, to reference a method included in a module, the standard syntax (\ie, \texttt{module.method}) can be replaced with these alternatives: \texttt{getattr(module, "method")}, \codeinline{module.\_\_getattribute\_\_("method")}, as well as \codeinline{module.\_\_dict\_\_["method"]}.
These obfuscation techniques can also be combined to further complicate analysis.
For instance, the payload \codeinline{"bash -i >& /dev/tcp/10.0.0.1/8080 0>&1"} can be equivalently rewritten as:
\texttt{getattr(\_\_import\_\_("os"), "system").\_\_call\_\_}\codeinline{("bash -i >& /dev/tcp/10.0.0.1/8080 0>&1")}.

\subsection{Adversarial Training}\label{sec:adversarial_training}
To improve the adversarial robustness of our solution, we leveraged AT~\cite{madry2018advtrain,biggio2018wild}.
The core idea is to generate and include adversarial examples (\ie, adversarial packages, in our case) during training, thereby enabling the model to withstand the corresponding evasive attack patterns at test time. 
In common machine learning applications, such as image recognition, AT leverages gradient-based attacks to craft adversarial examples in the feature space~\cite{madry2018advtrain}, as the corresponding optimization problem is end-to-end differentiable. 
However, for several cybersecurity domains, including malware detection~\cite{Demontis2019YesML,demetrio2021tifs}, phishing~\cite{Montaruli2023AdvPhishing} and web application firewalls~\cite{modsecadvlearn}, this is not directly applicable 
as the considered transformations are not end-to-end differentiable, given the presence of non-differentiable feature extraction steps.
Moreover, gradient-based attacks cannot be applied to machine learning models that are not differentiable, such as tree-based models that are widely used in the field of malicious package detection~\cite{Ladisa2023ACSAC,Ohm2022ARES}.
For these reasons, following prior works in other cybersecurity domains~\cite{modsecadvlearn,demetrio2021tifs,Montaruli2023AdvPhishing}, we adopted problem-space AT, which leverages
practical functionality-preserving transformations, combined with black-box optimization algorithms.

Specifically, we leveraged the query-efficient black-box optimizer introduced by Montaruli \etal~\cite{Montaruli2023AdvPhishing}, which relies on mutation-based fuzzing techniques. 
Its goal is to mutate the original malicious package by leveraging the adversarial transformations described in Section~\ref{sec:transformations} to minimize the confidence score returned by the machine learning model (\ie, the objective function of the optimization),
using an iterative approach consisting of consecutive mutation rounds.

To improve query efficiency, the transformations are categorized into single-round (SR) and multi-round (MR).
The former (SR) are applied only once to the input sample, while the latter (MR) are applied iteratively to the adversarial package generated in the previous round.
In our implementation, all data obfuscation transformations are implemented as SR transformations: for each type of IOC (IP, URL, system commands), we randomly select one transformation from encoding, compression, encryption, or binary array representations and apply it to all the related strings in the source code.
As for the static code transformations, they are implemented as SR transformations except for useless code injection (treated as a MR transformation) since the amount of code required to reduce the model's confidence score (\ie, to evade detection) is not known a priori.

\section{Experiments}
\label{sec:exp_results}
In this section we describe the experimental setup, the detectors and the datasets used in our tests.
Finally, we introduce the research questions we want to explore in our experiments.

\subsection{Detectors}\label{sec:detectors}
Several detectors have been proposed by the research community to detect supply-chain attacks, often using different approaches 
and relying on different features.
As a baseline, we will use GuardDog~\cite{datadog2022guarddog}, a tool for malicious package detection based on static rules.

Most of the solutions based on machine learning that have been proposed to date are not available, and
those that are, such as the one proposed by Ladisa \etal~\cite{Ladisa2023ACSAC},
do not include several key features that have been shown to be very effective in detecting malicious packages~\cite{zhang2023cerebro,huang2024donapi}.

Hence, to build our state-of-the-art detector (\texttt{SoA} hereinafter), we took the open-source tool released by Ladisa \etal~\cite{Ladisa2023ACSAC} as a starting point 
and extended its feature set to include missing key features to capture the presence of security-sensitive APIs and suspicious behaviors~\cite{zhang2023cerebro,huang2024donapi}. 

The complete feature set of our \texttt{SoA} detector is summarized in \autoref{table:features_new_detector}.
On top of those already provided by the tool, we added the following three classes (highlighted in green in \autoref{table:features_new_detector} and detailed below):
API-related, behavior-related and obfuscation-related (based on the adversarial transformations described in Section~\ref{sec:transformations}) features.

\begin{table}[t]
    \centering
    \begin{adjustbox}{max width=\columnwidth}
    \begin{tabular}{c c c c}
    \toprule
    \textbf{Category} & \textbf{Features} & \textbf{Type} & \textbf{Size} \\
    \toprule
    \multirow{4}{*}{\shortstack{Structural}}
    & Presence of installation hook(s) in \texttt{setup.py} &  B  & 1 \\
    \cline{2-4}
    & Count of lines (source and metadata)  &  N  &  2  \\
    \cline{2-4}
    & Count of words (source and metadata)  &  N  &  2  \\
    \cline{2-4}
    &  Count of files per selected extensions (.js, .md, \dots)  &   N  &   91  \\
    \midrule
    \cellcolor{green!40} API & \cellcolor{green!40} Count of security-related API & \cellcolor{green!40} N  & \cellcolor{green!40} 215  \\
    \midrule
    \cellcolor{green!40} Behavior &  \cellcolor{green!40} Count of suspicious behaviors &  \cellcolor{green!40} N  & \cellcolor{green!40} 5  \\
    \midrule
    \multirow{10}{*}{\shortstack{Obfuscation}}
    & \cellcolor{green!40} Count of adversarial patterns (source and metadata) &  \cellcolor{green!40}  N  &  \cellcolor{green!40} 12 \\
    \cline{2-4}
    & \makecell{Statistics (mean, std. deviation, 3rd quartile and max) \\ of Shannon entropy of strings (source and metadata)} &  N & 8 \\
    \cline{2-4}
    & \makecell{Statistics (mean, std. deviation, 3rd quartile and max) \\ of Shannon entropy of identifiers (source and metadata)} &  N & 8 \\
    \cline{2-4}
    & \makecell{Count of homogeneous and \\ heterogenous strings (source and metadata)} &  N & 4 \\
    \cline{2-4}
    & \makecell{Count of homogeneous and \\ heterogenous identifiers (source and metadata)} &  N & 4 \\
    \midrule
    \multirow{10}{*}{\shortstack{String}}
    & \makecell{Count of URLs (source and metadata)} &  N & 2 \\
    \cline{2-4}
    & \makecell{Count of IP addresses (source and metadata)} &  N & 2 \\
    \cline{2-4}
    & \makecell{Count of suspicious tokens in \\ strings (source and metadata)} &  N & 2 \\
    \cline{2-4}
    & \makecell{Count of base64 strings (source and metadata)} &  N & 2 \\
    \cline{2-4}
    & \makecell{Statistics (mean, std. deviation, 3rd quartile and max) \\ of ratio of square brackets per source code file size} &  N & 8 \\
    \cline{2-4}
    & \makecell{Statistics (mean, std. deviation, 3rd quartile and max) \\ of ratio of equal signs per source code file size} &  N & 8 \\
    \cline{2-4}
    & \makecell{Statistics (mean, std. deviation, 3rd quartile and max) \\ of ratio of plus signs per source code file size} &  N & 8 \\
    \bottomrule
    \end{tabular}
    \end{adjustbox}
    \caption{Summary of the features adopted by the \texttt{SoA} detector. N: numeric feature, B: boolean feature.
    Features in green are those added in this work.}
    \label{table:features_new_detector}
\end{table}

\myparagraphlb{API-related features.}
These features are designed to detect the usage of security-sensitive APIs that are commonly used to perform malicious actions, such as executing arbitrary code, downloading files, or exfiltrating sensitive information.
To this end, we performed a detailed analysis of the most common APIs used in malicious packages of the \emph{MalwareBench} dataset~\cite{malwarebench} and the current literature~\cite{zhang2023cerebro}.
We identified a set of 215 APIs that are divided into six categories:
\texttt{Network}, \texttt{Filesystem}, \texttt{Host Information}, \texttt{Code Execution}, \texttt{Command Execution} and \texttt{Encoding}.
For each API, we used one numeric feature representing its number of occurrences in the source code.
The complete list of APIs is shown in \autoref{tab:api} in Appendix~\ref{sec:appendix-new-features}.

\myparagraphlb{Behavior-related features.}
These features are designed to detect the presence of suspicious behaviors that are commonly associated with malicious packages.
To this end, we adopted the behaviors defined by Guo \etal~\cite{Guo2023EmpiricalStudyPyPI}, namely \emph{Remote Control}, \emph{Information Stealing},
\emph{Code Execution}, \emph{Command Execution} and \emph{Unauthorized File Operations}.
Each behavior consists of one or more sequences of security-sensitive APIs belonging to the categories defined above.
For instance, the \emph{Information Stealing} behavior is defined as a sequence of APIs belonging to the following categories: ([\texttt{Filesystem}], \texttt{Host Information}, [\texttt{Code Execution}], \texttt{Network}),
where the brackets indicate that the API is optional.

To detect the presence of these behaviors, we leveraged a two-step approach:
(i) security-sensitive APIs extraction and
(ii) behavior matching.
In the first step, for each file, we extracted the security-sensitive APIs from the Abstract Syntax Tree (AST) representation of the code by leveraging the \texttt{ast} module in Python.
In the second step, we replaced each API with the corresponding category and matched the obtained sequence of categories 
against the predefined behaviors, which are represented through battle-tested regular expressions.
The regular expressions are designed to be flexible, \ie, they allow for the presence of additional API categories in the sequence, as long as the order of the categories is preserved.
For instance, the \emph{Information Stealing} behavior can be detected in the (\texttt{Host Information}, \texttt{Command Execution}, \texttt{Network}) sequence, where there is a \texttt{Command Execution} API between \texttt{Host Information} and \texttt{Network} APIs.
Furthermore, we extensively tested the regular expressions on the \emph{MalwareBench} dataset by manually verifying that they are able to detect the behaviors of the malicious packages in the dataset.
For each behavior, we used one numeric feature to count the number of occurrences of the behavior in the package.
The complete list of behaviors and the related regular expressions are shown in \autoref{tab:behaviors} in Appendix~\ref{sec:appendix-additional-experiments}.

\myparagraphlb{Obfuscation-related features.}
These features are designed to detect the presence of various obfuscation patterns that are commonly used in malicious packages.
To this end, we leveraged a set of battle-tested regular expressions aimed at detecting
the main obfuscation techniques related to the proposed adversarial transformations (see Section~\ref{sec:transformations}),
including all the data obfuscation techniques (such as Base64, hexadecimal and binary encoding, as well as string splitting),
along with API obfuscation.
For each obfuscation technique, we used one numeric feature to count the number of occurrences of the obfuscation pattern in both the source code and metadata (\ie, \texttt{setup.py}).
For further implementation details, we refer the reader to Appendix~\ref{sec:appendix-new-features}.

\subsubsection{Model Training and Evaluation}
All experiments were conducted on an Ubuntu 22.04.6 LTS server equipped with an Intel Xeon Platinum 8160 CPU @ 2.10 GHz (64 cores) and 256 GB of RAM.

To train and evaluate our detectors, we leveraged all the tree-based models proposed by Ladisa \etal~\cite{Ladisa2023ACSAC}, namely Decision Tree, Random Forest~\cite{breiman2001rf} and XGBoost~\cite{xgboost}.
We focused only on these models for several reasons.
First, for a fair evaluation with their work.
Second, tree-based models are widely adopted in this area (e.g., by Huang \etal~\cite{huang2024donapi}) for their advantages: explainability and effectiveness at handling high-dimensional data, while maintaining good accuracy~\cite{molnar2022}.
Third, tree-based models ensure a better trade-off between performance and computational training cost compared to deep-learning solutions~\cite{tree_based_models_neurips}, which is very important in both our enterprise and PyPI scenarios.

All models were implemented using the following Python libraries:
\verb|scikit-learn| v1.5.0~\cite{sklearn} and \verb|xgboost| v2.1.0.
To train and tune the models' hyper-parameters, in line with current research~\cite{Trizna2024Nebula},
we performed a grid search based on a 5-fold cross-validation (CV) on the training set~\cite{raschka2018} to ensure a fair evaluation.
Since this resulted in five different detectors (for each tree-based model), all the results reported in the following are the mean values of the five detectors,
each one evaluated on the corresponding test set obtained by the 5-fold CV.

\subsection{Datasets}
For our experiments, we adopted two types of datasets.
First, in order to compare with the state-of-the-art and experiment with our detector, we employed \emph{MalwareBench} --
a recent dataset collected in fall 2023 by Zahan et al.~\cite{malwarebench}.

In addition, to perform real-world tests and measure how the \texttt{SoA} detector performed in the wild, we collected
a live stream of temporally-newer package releases from PyPI over two different time periods.

\myparagraphlb{MalwareBench.}
\emph{MalwareBench} is a state-of-the-art dataset that includes 3,190 unique malicious and 3,368 unique benign Python packages.
We use this dataset to evaluate the robustness of models against adversarial transformations, as well as the effectiveness of AT and the \texttt{SoA} features.
We adopted the same approach as Scano \etal~\cite{modsecadvlearn} for splitting the MalwareBench dataset to evaluate robustness and perform adversarial training.
As for the robustness evaluation, for each \texttt{test set} generated by the 5-fold CV, we create the corresponding \texttt{adversarial test set} (\texttt{test-adv}),
which contains the same benign samples as the original test set, but with the malicious samples generated from the original malicious samples by applying the proposed transformations.

As for AT, for each fold, we generate an \texttt{adversarial training set} by applying the adversarial transformations to the malicious samples in the training set.
Each \texttt{adversarial training set} is then merged with the corresponding training set to re-train the models.

Finally, to evaluate the robustness of the models re-trained with AT, we generate a new \texttt{adversarial test set} optimized on the re-trained models.

We would like to clarify that, although using the same transformations and optimizer,
the adversarial packages generated for building the \texttt{adversarial training sets} and \texttt{adversarial test sets} are different.
Indeed, they are \emph{independently optimized} against each target model at test time, resulting in the application of different, optimal transformation strategies.
Hence, the two sets are independent, ensuring an unbiased evaluation.

As for GuardDog, we evaluated its baseline detection capabilities on the \texttt{test set} and its adversarial robustness on
the \texttt{adversarial test set}, but we did not leverage AT since it is not a machine learning-based model.

\myparagraphlb{Real-world datasets.}
We built two real-world datasets by collecting all packages uploaded to PyPI over two different periods of time:
\texttt{live1} (for 80 days from 02/10/2024 to 21/12/2024) and \texttt{live2} (for 37 days from 31/03/2025 to 06/05/2025).
The \texttt{live1} dataset contains 48,712 unique packages and 122,398 releases, while the \texttt{live2} dataset contains 38,567 unique packages and 91,949 releases.
The datasets were built by leveraging the PyPI feeds for newly uploaded packages\footnote{\url{https://pypi.org/rss/packages.xml}} and releases\footnote{\url{https://pypi.org/rss/updates.xml}} in the specified time period.
We filtered out the packages without source code.
After each vetting period, we collected the ground truth labels of the packages by leveraging the Open Source Security Foundation (OpenSSF)~\cite{openssf_db_malware} and PyPI~\cite{pypi_db_malware} databases 
of malicious packages.
In addition, we carefully analyzed all the packages reported as malicious by the detectors to ensure that they were actually malicious or false positives.
The daily malicious packages found by the detectors were also reported to the PyPI maintainers.

The real-world datasets are used to perform several experiments.
First, to evaluate the impact of AT and the \texttt{SoA} features on the detection capabilities of the models in the wild over two different time periods.
Additionally, we leveraged the \texttt{live1} dataset to tune the number of adversarial packages to be used for AT and evaluate the detection capabilities of the models on obfuscated packages.
On the other hand, the \texttt{live2} dataset is used in our first case study to perform two experiments:
(i) to evaluate the impact of using \texttt{live1} to re-train the detectors, \ie, if using a greater variety of packages can improve the detection capabilities of the models,
(ii) to evaluate the performance of the final detector (based on the \texttt{SoA} features and trained using both \texttt{live1} and AT) in production for vetting temporally-newer packages uploaded to PyPI, 
when tuned with a very low FPR threshold (0.1\%).

\subsection{Research Questions}
\myparagraph{[RQ.1] Adversarial transformations} -- How effective are the proposed adversarial transformations 
in bypassing the evaluated detectors?

\vspace{0.1cm}
\myparagraph{[RQ.2] Adversarial robustness} -- To what extent does AT improve the adversarial robustness of the tested detectors?

\vspace{0.1cm}
\myparagraph{[RQ.3] Detection capabilities in the wild} -- How effective is the \texttt{SoA} detector based on AT in finding malicious packages in the wild?
Is it able to detect real (obfuscated) malicious packages that are undetected by the corresponding baseline?

\vspace{0.1cm}
\myparagraph{[RQ.4] Practical Deployment} -- 
How practical is the final detector when deployed in different production settings? In particular, how much effort is needed
to verify the daily alerts in two opposite case studies (PyPI deployment tuned at 0.1\% FPR and industrial deployment tuned at 10\% FPR)?

\vspace{0.1cm}
To answer these questions, we performed an extensive evaluation and discuss the results in the next two sections.

\section{Results}
\label{sec:results_baseline}

\myparagraphlb{Baseline Evaluation.}
\autoref{tab:results_detectors} shows the recall (\aka True Positive Rate or TPR), at 1\% FPR for the target detectors evaluated on the \textit{MalwareBench} dataset.
Moreover, we decided to use this operational point (1\% FPR) since it is a common practice in the literature~\cite{Trizna2024Nebula,Ponte2025SLIFER,Demontis2019YesML,Montaruli2023AdvPhishing,modsecadvlearn}, 
and it allowed us to perform a fair comparison among the detectors.
For completeness, in \autoref{fig:roc_detectors_all} in Appendix~\ref{sec:appendix-additional-experiments} 
we also provide the Receiver Operating Characteristic (ROC) curves of all the detectors.

\begin{table}[t]
    \centering
    \begin{adjustbox}{max width=\columnwidth}
    \begin{tabular}{c c c}
        \toprule
        \textbf{Detector} & \multicolumn{2}{c}{\textbf{MalwareBench Dataset}} \\
        \cmidrule(lr){2-3}
        &   \texttt{test} & \texttt{test-adv} \\
        \toprule
        GuardDog             &   6.63   &  1.00 \\
        \midrule
        Decision Tree  &  69.14  &  4.60 \\
        Random Forest  &  90.54  &  12.10 \\
        XGBoost        &  95.27  &  24.30 \\
        \midrule
        XGBoost AT     &   \textbf{95.64}  &  \textbf{86.81} \\
        \bottomrule
    \end{tabular}
    \end{adjustbox}
    \vspace{+0.1cm}
    \caption{Recall (TPR) at 1\% FPR of different detectors evaluated on the \texttt{test} (baseline performance) and \texttt{test-adv} (adversarial robustness) sets.}
    \label{tab:results_detectors}
\end{table}

We evaluated the performance of all detectors on the two variants of the dataset: \texttt{test} and \texttt{test-adv}.
The former is used to evaluate the baseline performance, while the latter to evaluate the adversarial robustness of the detectors.
The results highlight several important points.
Among the different models, XGBoost is the one with the best performance, with a 95.27\% detection rate at 1\% FPR. 
On the other end of the spectrum, GuardDog is the one with the worst performance,  detecting only 6.63\% of the malicious packages in the vanilla dataset.
This result, aligned with the findings of Vu \etal~\cite{Vu2023BadSnakes}, highlights that current solutions based on static rules do
not perform well at low FPR compared to machine learning-based approaches.

However, not surprisingly, all approaches performed poorly on adversarial samples.
Also in this case, GuardDog was the worst (1\% detection) and XGBoost the best (24.3\% detection).
This shows that the adversarial packages generated by our transformations
can easily bypass the current state-of-the-art detectors.

Based on these results, we use the best-performing XGBoost model for the rest of the experiments.

\vspace{0.2cm}
\takeawaysNoTitle{
    \textbf{[RQ.1]}   
The adversarial transformations proposed in our study are very effective at bypassing the current state-of-the-art detectors.
The detection rate of the best model decreased from 95\% to a mere \textbf{24\%} when tested on obfuscated packages.
}

\vspace{0.4cm}
\myparagraphlb{Adversarial Training.}
To apply AT, we first performed a preliminary evaluation to
select the number of adversarial packages to include in the model training.
To this end, we first sorted the adversarial packages in the \texttt{adversarial training set} based on the output score of the model,
and then we selected the top $k$ adversarial packages with the highest output score, where $k$
is the percentage of adversarial packages to be used for training the model.
We then evaluated the resulting models on the \texttt{live1} dataset and reported the performance in \autoref{tab:results_at_study}.
Our experiments show that the model performance improves by adding adversarial samples, but if too many (such as 100\%)
of them are added to the training set, the performance decreases.
Overall, we found that using 20\% of the adversarial packages provides the best recall and F1-score.
Hence, we used 20\% of the adversarial packages for training the detectors with AT in the following experiments.

The last line of Table~\ref{tab:results_detectors} shows the results of the XGBoost model trained with AT on the \texttt{MalwareBench} dataset.
We can clearly see that AT significantly improves the robustness of the XGBoost model against the proposed adversarial transformations 
while keeping the same performance on the baseline \texttt{test set}.

\begin{table}[!t]
    \centering
    \begin{adjustbox}{max width=\columnwidth}
    \begin{tabular}{c c c c c c c c c c}
        \toprule
        \textbf{Model} & \textbf{FP} & \textbf{TP} & \textbf{FN} & \textbf{TN} & \textbf{Acc} & \textbf{Prec} & \textbf{Rec} & \textbf{F1} \\
        \toprule
        \texttt{base}          &  1210  &  242  &  112  &  120832  &  98.92  &  16.67  &  68.36  &  26.80 \\
        \texttt{AT-10}  &  1219  &  223  &  131  &  120823  &  98.90  &  15.46  &  62.99  &  24.83 \\
        \textbf{\texttt{AT-20}}  &  \textbf{1210}  &  \textbf{246}  &  \textbf{108}  &  \textbf{120832}  &  \textbf{98.92}  &  \textbf{16.90}  &  \textbf{69.49}  &  \textbf{27.18} \\
        \texttt{AT-30}  &  1217  &  226  &  128  &  120825  &  98.90  &  15.66  &  63.84  &  25.15 \\
        \texttt{AT-40}  &  1212  &  231  &  123  &  120830  &  98.91  &  16.01  &  65.25  &  25.71 \\
        \texttt{AT-50}  &  1218  &  237  &  117  &  120824  &  98.91  &  16.29  &  66.95  &  26.20 \\
        \texttt{AT-60}  &  1215  &  226  &  128  &  120827  &  98.90  &  15.68  &  63.84  &  25.18 \\
        \texttt{AT-70}  &  1216  &  239  &  115  &  120826  &  98.91  &  16.43  &  67.51  &  26.42 \\
        \texttt{AT-80}  &  1220  &  227  &  127  &  120822  &  98.90  &  15.69  &  64.12  &  25.21 \\
        \texttt{AT-90}  &  1217  &  220  &  134  &  120825  &  98.90  &  15.31  &  62.15  &  24.57 \\
        \texttt{AT-100} &  1219  &  146  &  208  &  120823  &  98.83  &  10.70  &  41.24  &  16.99 \\
        \bottomrule
    \end{tabular}
    \end{adjustbox}
    \vspace{+0.1cm}
    \caption{Performance metrics at 1\% FPR of the XGBoost model based on the \texttt{SoA} features evaluated on the \texttt{live1} dataset.
    \texttt{base} represents the model trained on the main training set, while \texttt{AT-X} represents the model trained using AT with the specified percentage (\texttt{X}) of adversarial samples.
    The best results are highlighted in bold.}
    \label{tab:results_at_study}
\end{table}

\vspace{0.2cm}
\takeawaysNoTitle{
\textbf{[RQ.2]} AT significantly improves the robustness against the same set of transformations up to \textbf{2.5$\bm{\times}$ increase} over the baseline.
}

\begin{figure*}[t]
    \begin{tabular}{C{.5\textwidth}C{.5\textwidth}}
        \textbf{Obfuscated} & \textbf{Non-Obfuscated} \\
        \parbox{0.48\textwidth} {
            \begin{tikzpicture}
            \def\radius{1cm}
            \def\mycolorbox#1{\textcolor{#1}{\rule{2ex}{2ex}}}
            \colorlet{colori}{blue!100}
            \colorlet{colorii}{green!100}
            
            \coordinate (ceni);
            \coordinate[xshift=\radius] (cenii);
            
            \draw[fill=colori,fill opacity=0.5] (ceni) circle (\radius);
            \draw[fill=colorii,fill opacity=0.5] (cenii) circle (\radius);
            
            \node at ([xshift=2.5cm] current bounding box.east) 
            {
                
            \begin{tabular}{@{}ll@{\,\,}c@{}}
                & Total: & 66  \\
            \mycolorbox{colori!50} & \texttt{Base}:& 59 \\
            \mycolorbox{colorii!50} & \texttt{AT}:&  65 \\
            \end{tabular}
            };
            
            \node[xshift=-.5\radius] at (ceni) {$1$};
            \node[xshift=.5\radius] at (cenii) {$7$};
            \node[xshift=.4\radius] at (ceni) {$58$};
    
            \end{tikzpicture}
            \vspace{+0.2cm}
        } &
        \parbox{0.48\textwidth} {
        \begin{tikzpicture}
            \def\radius{1cm}
            \def\mycolorbox#1{\textcolor{#1}{\rule{2ex}{2ex}}}
            \colorlet{colori}{blue!100}
            \colorlet{colorii}{green!100}
            
            \coordinate (ceni);
            \coordinate[xshift=\radius] (cenii);
            
            \draw[fill=colori,fill opacity=0.5] (ceni) circle (\radius);
            \draw[fill=colorii,fill opacity=0.5] (cenii) circle (\radius);
            
            \node at ([xshift=2.5cm] current bounding box.east) 
            {
                
            \begin{tabular}{@{}ll@{\,\,}c@{}}
                & Total: & 188  \\
            \mycolorbox{colori!50} & \texttt{Base}:& 183 \\
            \mycolorbox{colorii!50} & \texttt{AT}:& 181 \\
            \end{tabular}
            };
            
            \node[xshift=-.5\radius] at (ceni) {$7$};
            \node[xshift=.5\radius] at (cenii) {$5$};
            \node[xshift=.4\radius] at (ceni) {$176$};
    
            \end{tikzpicture}
        } \\
    \end{tabular}
    \caption{Comparison between the baseline and the AT-based models on the \texttt{live1} dataset 
    in terms of detection of obfuscated (left) and non-obfuscated (right) samples.
    }
    \label{fig:results_obfuscation}
\end{figure*}
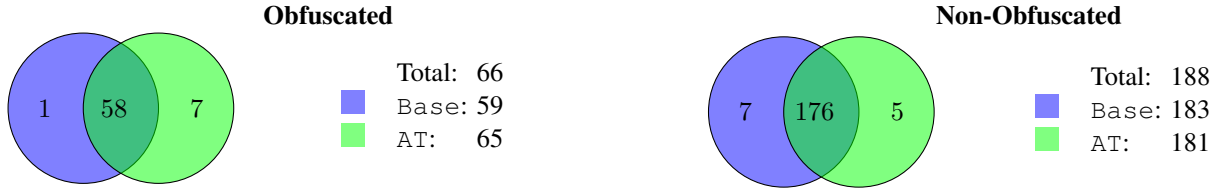

\begin{table*}[!t]
    \centering
    \begin{adjustbox}{max width=\textwidth}
        \begin{tabular}{c c c c c c c c c c c}
        \toprule
        \textbf{Model} &  \textbf{Training Dataset}  &  \textbf{FP} & \textbf{TP} & \textbf{FN} & \textbf{TN} & \textbf{Acc} & \textbf{Prec} & \textbf{Rec} & \textbf{F1} \\
        \toprule
        \multirow{2}{*}{XGBoost base}  &   \texttt{train}     &  798  &  147  &  69  &  90,810  &  99.06  &  15.56  &  68.06  &  25.32 \\
                                       &   \texttt{train + live1}     &  786  &  167  &  49  &  90,822  &  99.09  &  17.52  &  77.31  &  28.57 \\
        \midrule
        \multirow{2}{*}{XGBoost AT}  &  \texttt{train} &  792  &  153  &  63  &  90,816  &  99.07  &  16.19  &  70.83  &  26.36 \\
                                     &  \texttt{train + live1} &  \textbf{781}  &  \textbf{180}  &  \textbf{36}  &  \textbf{90,827}  &  \textbf{99.11}  &  \textbf{18.73}  &  \textbf{83.33}  &  \textbf{30.59} \\ 
        \bottomrule
    \end{tabular}
    \end{adjustbox}
    \vspace{+0.1cm}
    \caption{Performance metrics at 1\% FPR of the XGBoost models evaluated on the \texttt{live2} dataset.
    }
    \label{tab:results_live2}
\end{table*}

\vspace{0.4cm}
\noindent
\textbf{Real-World Experiments.}
\label{sec:results_realworld}
We now present the results of our evaluation on the \texttt{live1} dataset.
As explained before, this was conducted on a live PyPI feed by using the XGBoost model.
The results are reported in \autoref{tab:results_at_study}.

For this experiment we compared the XGBoost model trained with AT using 20\% of the adversarial packages (the best configuration -- see \autoref{tab:results_at_study}) 
and the corresponding baseline model (\texttt{base} -- trained without AT).

As a first observation, we can see that the two detectors have very similar performance,
with the one trained with AT performing slightly better overall
(+1\% detection rate). This small increment shows that while AT is important,
malicious packages observed in the wild today adopt the obfuscation techniques we described in the paper only to a limited extent.

This is confirmed by \autoref{fig:results_obfuscation}, which shows a detailed analysis of the malicious packages found by the detectors in the \texttt{live1} dataset.
Overall, the two models (baseline and AT-based) detected \underline{254 malicious packages}, of which 66 (26\%) showed some form of obfuscation based on the adversarial patterns introduced in this work (see Appendix~\ref{sec:appendix-obfuscation-features}).
Moreover, by further analyzing the obfuscated packages, we found that only 31 of them (12.2\% of the total) leverage more than one obfuscation technique.

Furthermore, it is interesting to observe that the AT-based model performs better on obfuscated packages,
finding 6 (+10.2\% increase\footnote{percentage increment w.r.t the baseline: $(65 - 59) / 59 \times 100 = 10.2\%$}) more obfuscated packages compared to the baseline model,
while on non-obfuscated packages the baseline model has a small edge (+2 package detected compared to the AT-based model).
This seems to suggest that AT may cause the model to overfit on obfuscated packages at the expense of non-obfuscated ones,
which are however more common in the wild.

It is also important to notice that both detectors perform worse on real-world data than on the \texttt{test} dataset -- with a
drop in detection rate at 1\% FPR from roughly 95\% to 69\%.
Nevertheless, such result is also in line with the findings of Zhang \etal~\cite{zhang2023cerebro}, 
who observed an even higher drop in performance when evaluating their detector on real-world data (precision decreased from roughly 95\% to 18.5\%).
This remarks the importance of evaluating the detectors on real-world data as current datasets might not be representative of the actual packages available on PyPI.

\vspace{0.2cm}
\takeawaysNoTitle{
\textbf{[RQ.3]}
Our experiments show that AT needs to be applied cautiously: on the one hand it makes the model more robust to obfuscations and allows to find 6 (+10\%) 
more obfuscated packages, but on the other hand it might negatively affect the detection of non-obfuscated packages.
}

\vspace{0.4cm}
\noindent
\textbf{Ablation study.}
\label{sec:ablation_study}
We also performed an ablation study to assess the impact of the new
features, such as the presence of security-sensitive APIs and obfuscation patterns, on the performance of the detector.

For space reasons, we only report here the key findings of our ablation study, while all the details are provided in Appendix~\ref{sec:appendix-ablation-study}.
Our experiments show that while the \texttt{SoA} features provide only a slight improvement in detection capabilities on the baseline dataset (\texttt{MalwareBench}),
they are crucial for enhancing the generalization capabilities of the models on novel samples from the real-world datasets (\texttt{live1} and \texttt{live2}).
This suggests that new features are essential for better capturing the malicious behaviors of real-world malware samples and, consequently,
improving the detection capabilities of the models.
\section{Case Studies}
\label{sec:results_production}

In this section, we present two case studies that show how the same detector can be deployed
in different settings, by tuning its operating point to either maximize its detection rate or 
to minimize the number of false alarms.

For this purpose, as depicted in \autoref{tab:results_live2}, we experimented with our XGBoost detector in four different configurations:
baseline and AT-based models trained on both the \texttt{train} and \texttt{live1} datasets (\texttt{train + live1}), and \texttt{train} dataset only.
The results confirm that the AT-based models perform better, and that using the larger and more realistic \texttt{live1} dataset in addition to the vanilla \texttt{train} dataset (based on MalwareBench)
results in stronger models with higher precision and recall.
This underlines that current state-of-the-art datasets, such as \texttt{MalwareBench}, are not fully representative of the real-world
distribution of packages, and that using a more realistic dataset can improve the detection capabilities of the models, especially when using the detector in production for vetting PyPI packages.
For this reason, for both case studies detailed in the following, we decided to deploy the final detector using the AT-based XGBoost model trained on (\texttt{train + live1}).

To avoid potential false negatives in \texttt{live1}, we selected all the packages with a SourceRank score\footnote{\url{https://docs.libraries.io/overview.html\#sourcerank}} 
(a popular ranking metric for open source packages developed by Libraries.io~\cite{Sun2023SoftwareProvenance,Saini2020SoftwareMetrics}) of at least 8.
We chose this threshold since it corresponds to the mean and median values of the benign packages in \texttt{live1}.
Finally, we reported the results in \autoref{tab:results_live2} at 1\% FPR to be consistent with the previous experiments,
while for vetting the packages in production our final detector was tuned at 0.1\% FPR.

\begin{table}[t]
    \centering
    \begin{adjustbox}{max width=\columnwidth}
        \begin{tabular}{c c c c c c c c c c c}
        \toprule
        \textbf{Max FPR} &  \textbf{FP} & \textbf{TP} & \textbf{FN} & \textbf{TN} & \textbf{Acc} & \textbf{Prec} & \textbf{Rec} & \textbf{F1} \\
        \toprule
        30\%  &  27478  &  208  &  8   &  64130  &  70.14  &  0.75   &  96.30   &  1.49 \\
        10\%  &  9152   &  196  &  20  &  82456  &  90.03  &  2.10   &  90.74   &  4.10 \\
        1\%   &  781    &  180  &  36  &  90827  &  99.11  &  18.73  &  83.33   &  30.59 \\
        0.1\% &  81     &  92   &  124 &  91527  &  99.78  &  53.18  &  42.59   &  47.30 \\
        0.05\% &  35     &  77  &  139 &  91573  &  99.81  &  68.75  &  35.65   &  46.95 \\
        \bottomrule
    \end{tabular}
    \end{adjustbox}
    \vspace{+0.1cm}
    \caption{Performance metrics at different FPR thresholds of the final detector evaluated on the \texttt{live2} dataset.
    }
    \label{tab:results_final_detector_live}
\end{table}

\subsection{PyPI}

In the first case study, we consider the scenario of a PyPI maintainer who wants to pre-screen every new release to filter out possible malicious packages.
In this case, the problem is that the total number of new releases published every day is very high, and the maintainers
have very little time to invest in this task ($\sim$20 minutes per week~\cite{Vu2023BadSnakes}).
Indeed, as reported in the interview conducted by Vu \etal~\cite{Vu2023BadSnakes} to the PyPI maintainers,
in 2020 PyPI introduced a malware scanning project, but it was discontinued two years later due to the overwhelming number of false positives.
Hence, it is of paramount importance to design a solution that generates very few false alarms.

\autoref{tab:results_final_detector_live} reports the results of this case study at different configuration points, ranging from 0.05\% FPR to 30\% FPR.
We would like to clarify that, even though during the vetting of the packages we used a threshold of 0.1\% FPR,
we report the performance of the final detector at different FPR thresholds for completeness and to show its flexibility in production.
When tuned at 0.1\% FPR, the final detector was able to detect 92 malicious packages (\ie, 42.59\% of the total) in the \texttt{live2} dataset with only 81 false positives.

From the point of view of a PyPI maintainer, this is a very promising result.
Indeed, considering that the \texttt{live2} dataset contains packages collected over a period of 37 days, 
this means that the final detector is able to detect,
on average, 2.48 malicious package per day with just 2.18 false alarms.
In other words, in this configuration the model only raises $\sim$4 alerts per day, and half of them correspond to real malicious packages.
These results perfectly align with the time budget of 20 minutes per week to review the false alarms suggested by the PyPI maintainers interviewed by Vu \etal~\cite{Vu2023BadSnakes}.
Indeed, assuming the time budget of $\sim$1 minute to triage a single alert suggested by Vu \etal~\cite{Vu2023BadSnakes}, we would have $\sim$15 minutes\footnote{computed as 2.18 (FPs) $\times$ 7 (days) $\times$ 1 (min) = 15.26 minutes} per week to review false positives.

Finally, we remark that the proposed detector also satisfies the other requirements highlighted by Vu \etal~\cite{Vu2023BadSnakes} in the context of the PyPI ecosystem, 
namely the low effort to use and maintain, and the real-time detection of malicious packages.
Indeed, our detector can be easily trained and deployed in parallel, can be easily tuned at different thresholds, and can scan packages in real-time (hundreds of milliseconds per package on average), making it a perfect solution for being integrated in the PyPI ecosystem.

\begin{table}[!t]
    \centering
    \begin{adjustbox}{max width=\columnwidth}
    \begin{tabular}{c c c}
    \toprule
    \textbf{Campaign} & \textbf{Count} & \textbf{Num Packages} \\
    \toprule
    Stealer & 14 & 61 \\
    PoC & 3 & 13 \\
    Dropper & 3 & 8 \\
    Trojan & 2 & 5 \\
    \bottomrule
    \end{tabular}
    \end{adjustbox}
    \caption{Campaigns of the malicious packages detected by the final detector in the \texttt{live2} dataset.}
    \label{table:campaigns_live2}
\end{table}

\begin{table}[!t]
    \centering
    \begin{adjustbox}{max width=\columnwidth}
    \begin{tabular}{c c}
    \toprule
    \textbf{Obfuscation} & \textbf{Num Packages} \\
    \toprule
    BaseXX Encoding  &  40 \\
    Hex Encoding  &  6  \\
    Binary Arrays &  1  \\
    Data Reordering  &  2  \\
    API Name Obfuscation  &  1 \\
    \bottomrule
    \end{tabular}
    \end{adjustbox}
    \caption{Obfuscation techniques of the malicious packages detected by the final detector in the \texttt{live2} dataset.}
    \label{table:obfuscation_live2}
\end{table}

\begin{table}[!t]
    \centering
    \begin{adjustbox}{max width=\columnwidth}
    \begin{tabular}{c c}
    \toprule
    \textbf{Obfuscation} & \textbf{Num Packages} \\
    \toprule
    \codeinline{other source code} &  38 \\
    \codeinline{setup.py} &  35 \\
    \codeinline{__init__.py} &  19 \\
    \bottomrule
    \end{tabular}
    \end{adjustbox}
    \caption{Location of the malicious code in the packages detected by the final detector in the \texttt{live2} dataset.
    The \codeinline{other} category includes all the files that are not \codeinline{setup.py} or \codeinline{__init__.py} files, such as \codeinline{main.py}.
    }
    \label{table:malicious_code_location_live2}
\end{table}

\newpage
\myparagraphlb{Malware behaviors and campaigns.}
We performed a detailed analysis of the \underline{92 malware packages detected} by the final detector in the \texttt{live2} dataset by studying the malware behaviors and campaigns of the detected packages.
As for campaign attribution, we manually grouped all the packages that share the same (or almost the same) malicious code into a single campaign.
Moreover, by analyzing the temporal distribution of the packages, we found that packages in the same campaign are usually uploaded to PyPI in a short time span (same day or within a few days).

As shown in \autoref{table:campaigns_live2}, we found 22 campaigns in total, which are distributed as follows: most of the campaigns (14 out of 22) are stealer packages (61 in total), followed by 3 PoC campaigns (13 packages), 3 dropper campaigns (8 packages), and 2 trojan campaigns with backdoor functionality (5 packages).
This result is in line with the findings of current research~\cite{rl_pypi_2024,fortinet_pypi_2024,zahan2024ShiftingTheLens}, which show an increasing trend of malicious packages with stealer behaviors.

Packages in the stealer campaigns aim to steal sensitive information from the victim's machine, such as passwords, browser cookies, cryptocurrency wallets, which are generally exfiltrated to a remote server, a Telegram bot, or a Discord channel.
We found some PoC malicious samples that connect to suspicious URLs and exfiltrate some basic information about the machine such as hostname and operating system (OS) version.
Packages in the dropper campaigns target Windows machines and aim to download and execute malicious binaries file from a remote server,
while the two trojan campaigns install a backdoor on the victim's machine.

Additionally, we analyzed the presence of obfuscation in the detected packages (see \autoref{table:obfuscation_live2}), based on the obfuscation features used in this study (see Section~\ref{sec:detectors}).
We found that 40 out of 92 packages (43.4\%) use at least one form of obfuscation techniques to hide the malicious code.
Among the obfuscated packages, all of them use at least one BaseXX (\eg, Base64, Base32, \dots) encoding, six of them leverage hexadecimal encoding,
while only two packages using data reordering (\eg, splitting strings in multiple chunks) and only one package uses binary arrays to obfuscate strings.
Moreover, we found just one package that uses a simple form of API obfuscation to import a module (\ie, \texttt{\_\_import\_\_("os")} instead of \texttt{import os}).

Finally, we analyzed the location of the malicious code in the packages (see \autoref{table:malicious_code_location_live2}), and found a very interesting result:
unlike the results reported by Guo \etal~\cite{Guo2023EmpiricalStudyPyPI},
who found that 68.6\% of the analyzed malicious packages contained the malicious code in the \codeinline{setup.py} file,
we instead found that the majority of them (38 out of 92, \ie, 41.3\%) contain the malicious code in other source files (\eg, \codeinline{main.py}) 
different from \codeinline{setup.py} or \codeinline{__init__.py},
while only 35 packages (38\% of the total) contain the malicious code in the \codeinline{setup.py} file,
and the remaining 19 packages (20.7\% of the total) include the malicious code in \codeinline{__init__.py}.

This result remarks that, even if the \codeinline{setup.py} file is still quite popular for placing the malicious code,
mainly because it is the first file that is automatically executed when a package is installed~\cite{Ladisa2023SCORED,Guo2023EmpiricalStudyPyPI},
\codeinline{__init__.py} and other source code files are becoming more and more popular among attackers to place the malicious code.
This may be because the \codeinline{setup.py} file is monitored by security researchers and tools such as GuardDog~\cite{Liang2023NeedleHaystack,Vu2023BadSnakes,datadog2022guarddog}, 
hence attackers might be trying to evade detection by placing the malicious code in less monitored files.

\begin{table}[!t]
    \centering
    \begin{adjustbox}{max width=\columnwidth}
        \begin{tabular}{c c c c}
        \toprule
        \textbf{Max FPR} &  \textbf{FP} & \textbf{TN} & \textbf{Acc}  \\
        \toprule
        30\%  &  130  &  1466  &  81.45 \\
        10\%  &  46   &  1550  &  97.12 \\
        1\%   &  2    &  1594  &  99.87 \\
        0.1\% &  0    &  1596  &  100.00 \\
        \bottomrule
    \end{tabular}
    \end{adjustbox}
    \vspace{+0.1cm}
    \caption{
    Performance metrics at different FPR thresholds of the final detector evaluated in the industrial case study.
    }
    \label{tab:results_final_detector_industrial}
\end{table}

\subsection{Industrial Scenario}
For the second case study we collaborated with a large multinational software company,
which provided us with a list of 584 Python dependencies used in some of its products.
We collected all the releases of the dependencies from PyPI during the same time period (37 days) of the first case study, obtaining a total of 1,596 packages.
In this case, it is more important to detect malicious packages even if this requires investing more time spent on validating false alerts.
For this reason, we configured our detector at 90\% detection rate,
which, according to Table~\ref{tab:results_final_detector_live}, corresponds to a 10\% FPR on the \texttt{live2} dataset.

For completeness, in this case study we also report the performance of the final detector at
different thresholds in \autoref{tab:results_final_detector_industrial}. 
Note that since there were no supply-chain attacks on any of the packages used by the company during our 
experiment, we cannot compute the detection rate.  

The results show that, when tuned at 10\% FPR, the final detector achieves 97.12\% accuracy.
This corresponds to 46 false alerts, an average of 1.24 false positives per day. 
This means that a security engineer has to spend only a few minutes per day to review false positives,
which is a very reasonable time budget for an industrial scenario.

At this scale, it would be possible to adopt an even more aggressive setup,
for instance by tuning the classifier at the [30\% FPR - 96\% TPR] point.
This would further increase the chances of detecting more attacks, while increasing
the number of false alarms to a still-manageable 3.5 per day (\ie, 130 in total over 37 days).

\vspace{0.2cm}
\takeawaysNoTitle{
    [\textbf{RQ.4}]
    When deployed in production our final detector demonstrated competitive performance and \textbf{very low effort to review false alarms (few minutes per day)} 
    for both PyPI maintainers (2.48 TPs and 2.18 FPs per day at 0.1\% FPR) and enterprise security teams (97.12\% accuracy and 1.24 FPs per day at 10\% FPR).
}

\section{Related Work}\label{sec:related_work}
In this section, we summarize the main solutions proposed to detect malicious packages in the PyPI ecosystem, focusing on their detection techniques and limitations.
Ladisa \etal~\cite{Ladisa2023ACSAC} proposed a cross-language malicious package detector that leverages static features and machine learning techniques to identify malicious packages in both the PyPI and NPM ecosystems.
They conducted a 10-day vetting campaign on PyPI and NPM, achieving a precision of 4.4\%.
Its main limitation is the lack of features representing the packages' behavior, such as API calls.
In our work, we overcome this limitation by extending their feature set with additional features that count security-relevant APIs, suspicious behaviors, and new obfuscation patterns.
Thanks to the extended \texttt{SoA} feature set and AT, we increased the precision to 18.73\% (3.25$\times$ higher than their solution).

Zhang \etal~\cite{zhang2023cerebro} proposed CEREBRO, a cross-language detector that leverages language models fine-tuned on malicious behavior sequences extracted from a package's source code.
They performed a real-world evaluation on PyPI over a seven-month period and achieved a precision of 18.2\%.
Compared to this approach, we achieved slightly better precision (\ie, 18.73\%) and, more importantly, our approach is significantly more computationally efficient,
as it does not require fine-tuning or inference with a language model, thus making our detector more suitable for real-time detection.

Liang \etal~\cite{Liang2023NeedleHaystack} proposed \textsc{MPHunter}, a solution that leverages clustering techniques to identify malicious packages.
The main limitation of this work is that their approach only analyzes the \texttt{setup.py} metadata file, while malicious code can reside in other files within the package.
Indeed, as highlighted in our analysis, the majority of the malicious packages identified in the \texttt{live2} dataset included malicious code in other source files (\eg, \codeinline{__init__.py}).

Current research also encompasses multiple solutions that employ traditional approaches that rely on static/dynamic analysis and rule matching to detect malicious packages.
For instance, Li \etal~\cite{malwukong} proposed \textsc{MalWuKong}, which leverages a combination of in-depth static analysis,
metadata information, and rule matching (using YARA and CodeQL).
They conducted a live evaluation on PyPI but did not report any precision or recall metrics.
Moreover, their solution does not seem suitable for real-time vetting, as it requires deep static analysis of the package's source code.
Recently, Zheng \etal~\cite{oscar_ase24} proposed \textsc{OSCAR}, a solution based on dynamic analysis that employs fuzz testing on exported functions and classes, as well as behavior monitoring via tailored API hooking.
However, this work also lacks precision and recall metrics in its real-world evaluation and, similarly to \textsc{MalWuKong}, it is unsuitable for real-time PyPI vetting due to the complexity of dynamic analysis.

Finally, some research works have explored the use of Large Language Models (LLMs) for detecting malicious packages 
For instance, Zahan \etal~\cite{zahanllmnpm2025} evaluated for the first time the effectiveness of LLMs in detecting malicious NPM packages, while 
Ibiyo \etal~\cite{ibiyo2025rag} evaluated the use of LLMs enhanced with Retrieval-Augmented Generation (RAG) techniques to detect malicious PyPI packages.

Overall, none of the existing works have comprehensively evaluated the adversarial robustness of their detectors against evasion attacks, nor have they investigated the effectiveness of adversarial training, especially in a real-world setting.
Also, none of the current solutions are designed to be customizable for different actors in the software supply chain, such as package maintainers or enterprise security teams.
By contrast, our approach addresses all these gaps by providing a robust and flexible solution that can be seamlessly integrated into both PyPI and enterprise ecosystems.
\section{Conclusions}
\label{sec:conclusions}
Securing the software supply chain nowadays requires robust and adaptable solutions that address the diverse needs of various stakeholders, from package maintainers to enterprise security teams.
To this end, we propose a flexible and effective detector for malicious PyPI packages that can be seamlessly integrated into both public and enterprise ecosystems.
We present the first study that thoroughly evaluates the robustness of a malicious package detector against adversarial code transformations and evaluates the impact of AT in this context.
Our experiments show the double-edged sword effect of adversarial training: while it significantly improves robustness against the proposed adversarial transformations by 2.5$\times$ compared to the state-of-the-art and 
increases the detection rate of newly obfuscated malicious packages by 10\%, it also leads to a small drop (-1\%) in performance on non-obfuscated packages.

Finally, we demonstrate its adaptability to different operational needs: from PyPI maintainers requiring very low FPR (2.5 malicious packages detected daily on average vs.\ 2.18 false positives per day at 0.1\% FPR),
to enterprise settings with higher FPR requirements (1.24 false positives per day at 10\% FPR).
Hence, our detector can be tailored to the requirements of different actors in the software supply chain, ensuring a balance between security and usability.

Overall, our vetting campaigns have identified and reported to the community 346 malicious packages.

We foresee several future research directions to further improve our solution.
First, we plan to evaluate our detector on other ecosystems, such as NPM, which has seen over 540,000 malicious packages in recent years~\cite{sonatype2024}.
Indeed, our approach can be easily adapted to other ecosystems by tailoring the adversarial transformations to the respective programming languages.
Second, even though we are aware of more advanced techniques based on deep learning and LLMs and plan to explore them in the future,
they would require more computational resources for training and inference compared to our solution, which may not be suitable for real-time detection.
Finally, we aim to incorporate new features based on dynamic analysis to design a new hybrid solution that leverages both static and dynamic techniques to enhance the detection capabilities.
\appendices

\begin{table*}[!t]
    \centering
    \begin{adjustbox}{width=0.8\textwidth}
    \begin{tabular}{c c c}
    \toprule
    \textbf{Category} & \textbf{Module} & \textbf{API} \\
    \toprule
    \multirow{15}{*}{\shortstack{Network}}
    & \texttt{requests} & \makecell{\texttt{get}, \texttt{post}, \texttt{put}, \texttt{patch}, \texttt{delete}, \texttt{request}, \texttt{Session}} \\
    \cline{2-3}
    & \texttt{socket} & \makecell{\texttt{socket}, \texttt{close}, \texttt{create\_connection}, \\ \texttt{create\_server}, \texttt{dup}} \\
    \cline{2-3}
    & \texttt{socket.socket} & \makecell{\texttt{bind}, \texttt{listen}, \texttt{accept}, \texttt{connect}, \texttt{connect\_ex}, \\ \texttt{close}, \texttt{detach}, \texttt{dup}, \texttt{shutdown}} \\
    \cline{2-3}
    & \texttt{webhook} & \texttt{send} \\
    \cline{2-3}
    & \texttt{Webhook} & \texttt{from\_url} \\
    \cline{2-3}
    & \texttt{aiohttp} & \texttt{request} \\
    \cline{2-3}
    & \texttt{aiohttp.ClientSession} & \texttt{get}, \texttt{post}, \texttt{put}, \texttt{patch}, \texttt{delete}, \texttt{ws\_connect} \\
    \cline{2-3}
    & \makecell{\texttt{http.client.} \\ \texttt{HTTP(S)Connection}} & \texttt{request}, \texttt{getresponse}, \texttt{putrequest}, \texttt{connect}, \texttt{close} \\
    \cline{2-3}
    & \texttt{urllib.request} & \texttt{urlopen}, \texttt{urlretrieve}, \texttt{Request} \\
    \cline{2-3}
    & \texttt{urllib3} & \texttt{request} \\
    \cline{2-3}
    & \makecell{\texttt{urllib3.connection.} \\ \texttt{HTTP(S)Connection}} & \makecell{\texttt{connect}, \texttt{request}, \\ \texttt{request\_chunked}, \texttt{getresponse}, \texttt{close}} \\
    \midrule
    \multirow{8}{*}{\shortstack{Filesystem}}
    & \texttt{os} & \makecell{\texttt{open}, \texttt{remove}, \texttt{rename}, \texttt{replace}, \texttt{truncate}, \\ \texttt{stat}, \texttt{lstat}, \texttt{fstat}, \texttt{chown}, \texttt{chmod}, \\ 
                              \texttt{lchmod}, \texttt{lchown}, \texttt{link}, \texttt{symlink}, \texttt{readlink}, \texttt{realpath}, \\ \texttt{unlink}, \texttt{rmdir}, \texttt{mkdir}, \texttt{makedirs}, \texttt{removedirs}, \\ 
                              \texttt{walk}, \texttt{listdir}, \texttt{chdir}, \texttt{fchdir}, \texttt{access}, \\ \texttt{startfile}, \texttt{mkfifo}, \texttt{mknod}, \texttt{pathconf}, \texttt{fpathconf} \\ 
                              \texttt{statvfs}, \texttt{fstatvfs}, \texttt{fsync}, \texttt{fdatasync}, \texttt{sync}, \texttt{fsync\_range}} \\
    \cline{2-3}
    & \texttt{shutil} & \makecell{\texttt{copyfile}, \texttt{copyfileobj}, \texttt{copy}, \texttt{copy2}, \\ 
                              \texttt{copytree}, \texttt{rmtree}, \texttt{move}} \\
    \cline{2-3}
    & \texttt{io.BufferedReader} & \texttt{read}, \texttt{read1}, \texttt{readinto}, \texttt{readinto1} \\
    \cline{2-3}
    & \texttt{io.BufferedWriter} & \texttt{write} \\
    \cline{2-3}
    & \texttt{\texttt{builtins}}  &  \texttt{open}, \texttt{read}, \texttt{write}, \texttt{readline}, \texttt{readlines}, \texttt{writelines} \\                          
    \midrule
    \multirow{14}{*}{\shortstack{Host Information}}
    & \texttt{getpass} & \texttt{getuser}, \texttt{getpass} \\
    \cline{2-3}
    & \texttt{socket} & \makecell{\texttt{gethostname}, \texttt{getpeername}, \\ \texttt{gethostbyname}, \texttt{getfqdn}} \\
    \cline{2-3}
    & \texttt{platform} & \makecell{\texttt{node}, \texttt{system}, \texttt{release}, \\ \texttt{version}, \texttt{machine}, \texttt{processor}, \\
                                    \texttt{architecture}, \texttt{platform}, \texttt{uname}, \\ \texttt{linux\_distribution}, \texttt{mac\_ver}, \texttt{win32\_ver}} \\
    \cline{2-3}
    & \texttt{winreg} & \makecell{\texttt{CreateKey}, \texttt{CreateKeyEx}, \texttt{ConnectRegistry}, \\ 
                                    \texttt{DeleteKey}, \texttt{DeleteKeyEx}, \texttt{DeleteValue}, \\ 
                                    \texttt{EnumKey}, \texttt{EnumValue}, \texttt{LoadKey}, \\ 
                                    \texttt{OpenKey}, \texttt{OpenKeyEx}, \texttt{QueryValue}, \\ 
                                    \texttt{QueryValueEx}, \texttt{SaveKey}, \texttt{SetValue}, \\ 
                                    \texttt{SetValueEx}, \texttt{QueryInfoKey}} \\
    \midrule
    \multirow{1}{*}{\shortstack{Code Execution}}
    & \texttt{builtins} & \texttt{exec}, \texttt{eval} \\
    \midrule
    \multirow{5}{*}{\shortstack{Command Execution}}
    & \texttt{subprocess} & \makecell{\texttt{getoutput}, \texttt{call}, \texttt{check\_output},
                                    \texttt{run}, \texttt{Popen}, \texttt{check\_call}} \\
    \cline{2-3}
    & \texttt{pty} & \texttt{fork}, \texttt{openpty}, \texttt{spawn} \\
    \cline{2-3}
    & \texttt{os} & \makecell{\texttt{popen}, \texttt{system}, \texttt{posix\_spawn}, \texttt{posix\_spawnp}, \\
                              \texttt{getenv}, \texttt{chmod}, \texttt{dup2}, \texttt{startfile},
                              \texttt{exec}$\ast$\footnote{We use \texttt{exec}$\ast$ to refer to all the \texttt{exec} functions, such as \texttt{execv}, \texttt{execve}, \texttt{execvp} \dots},
                              \texttt{spawn}$\ast$\footnote{We use \texttt{spawn}$\ast$ to refer to all the \texttt{spawn} functions, such as \texttt{spawnl}, \texttt{spawnle}, \texttt{spawnlp} \dots}} \\
    \midrule
    \multirow{10}{*}{\shortstack{Encoding}}
    & \texttt{base64} & \makecell{\texttt{b64encode}, \texttt{b64decode}, \\ 
                                    \texttt{urlsafe\_b64encode}, \texttt{urlsafe\_b64decode}, \\ 
                                    \texttt{standard\_b64encode}, \texttt{standard\_b64decode}, \\
                                    \texttt{b32encode}, \texttt{b32decode}, \texttt{b16encode}, \texttt{b16decode}, \\
                                    \texttt{b85encode}, \texttt{b85decode}, \texttt{encode}, \texttt{decode}} \\
    \cline{2-3}
    & \texttt{hashlib} & \makecell{\texttt{md5}, \texttt{sha1}, \texttt{sha224}, \texttt{sha256}, \texttt{sha384}, \texttt{sha512}} \\
    \cline{2-3}
    & \texttt{bytearray} & \makecell{\texttt{fromhex}, \texttt{hex}} \\
    \cline{2-3}
    & \texttt{zlib} & \makecell{\texttt{compress}, \texttt{decompress}} \\
    \cline{2-3}
    & \texttt{gzip} & \makecell{\texttt{compress}, \texttt{decompress}} \\
    \cline{2-3}
    & \texttt{lzma} & \makecell{\texttt{compress}, \texttt{decompress}} \\
    \cline{2-3}
    & \texttt{marshal} & \makecell{\texttt{load}, \texttt{loads}} \\
    \cline{2-3}
    & \texttt{\_\_pyarmor\_\_} & \makecell{} \\
    \bottomrule
    \end{tabular}
    \end{adjustbox}
    \caption{List of security-sensitive APIs and their categories.}
    \label{tab:api}
\end{table*}

\begin{table*}[!t]
    \centering
    \begin{adjustbox}{width=0.8\textwidth}
    \begin{tabular}{c c}
    \toprule
    \textbf{Behavior} & \textbf{Regular Expression} \\
    \toprule
    Remote Control & \codeinline{NETWORK_(NETWORK_)?\\w*(ENCODING_)?\\w*CMDEXEC} \\
    \makecell{Information \\ Stealing} & \makecell{\codeinline{FILESYSTEM_(HOSTINFO_|ENCODING_)?\\w*NETWORK|} \\ \codeinline{HOSTINFO_(FILESYSTEM_|ENCODING_)?\\w*NETWORK}} \\
    Code Execution & \codeinline{NETWORK_(ENCODING_)?\\w*CEXEC|ENCODING_\\w*CEXEC|CEXEC} \\
    \makecell{Command \\ Execution} & \makecell{\codeinline{CMDEXEC_(ENCODING_)?\\w*NETWORK|ENCODING_\\w*CMDEXEC_(ENCODING_)?\\w*NETWORK|} \\ \codeinline{CMDEXEC_\\w*ENCODING|ENCODING_\\w*CMDEXEC_\\w*ENCODING|ENCODING_\\w*CMDEXEC}} \\
    \makecell{Unauthorized File \\ Operations} &  \makecell{\codeinline{NETWORK_\\w*FILESYSTEM_\\w*CMDEXEC_\\w*FILESYSTEM|NETWORK_\\w*FILESYSTEM_\\w*} \\ \codeinline{CMDEXEC|FILESYSTEM_\\w*CMDEXEC_\\w*FILESYSTEM|FILESYSTEM_\\w*CMDEXEC}} \\
    \bottomrule
    \end{tabular}
    \end{adjustbox}
    \caption{Behaviors adopted in this work and related regular expressions.}
    \label{tab:behaviors}
\end{table*}

\section{SoA Features}\label{sec:appendix-new-features}

\subsection{API-related Features}
In \autoref{tab:api} we provide the list of security-sensitive APIs that we used to extract the related features.
The APIs are grouped into categories based on their functionality,
namely \texttt{Network}, \texttt{Filesystem}, \texttt{Host Information}, \texttt{Code Execution}, \texttt{Command Execution} and \texttt{Encoding}.

\subsection{Behavior-related Features}
\autoref{tab:behaviors} shows the behaviors adopted in this work and the related regular expressions.

\subsection{Obfuscation-related Features}\label{sec:appendix-obfuscation-features}
In this work we leverage several features based on our study about adversarial packages to detect the main obfuscation techniques commonly used in malicious packages.
To this end, we designed several regular expressions to match the presence of the following obfuscation techniques:
\texttt{baseXX} encoding (including  Base64, Base32, Base16 and Base85 encoding schemes), hexadecimal encoding, binary array encoding, string splitting, XOR-based obfuscation and API obfuscation.
As for the \texttt{baseXX} encoding, we use several regular expressions to detect the presence of related APIs, such as \texttt{b64decode()}, \texttt{b32decode()}, \texttt{b16decode()} and \texttt{b85decode()}.
The hexadecimal encoding is detected by matching the presence of the \texttt{bytes.fromhex()} and \texttt{hex()} methods, as well as the presence of hex-encoded strings.
The binary array encoding is detected by matching the presence of the \texttt{bytes()} and \texttt{bytearray()} methods as well as the presence of bytearray-encoded strings.
The string splitting obfuscation is detected by matching the presence of string concatenation based on the \texttt{+} operator and \texttt{join()} method.
The XOR-based obfuscation is detected by matching the presence of the \texttt{\^} operator used along with the \texttt{xor()} and \texttt{ord()} methods.
Finally, the API obfuscation is detected by matching the obfuscation patterns detailed in Section~\ref{sec:transformations} and summarized in \autoref{fig:api_obfuscation}.

\section{Additional results}\label{sec:appendix-additional-experiments}
\begin{figure}[!t]
    \centering
    \includegraphics[width=0.95\columnwidth]{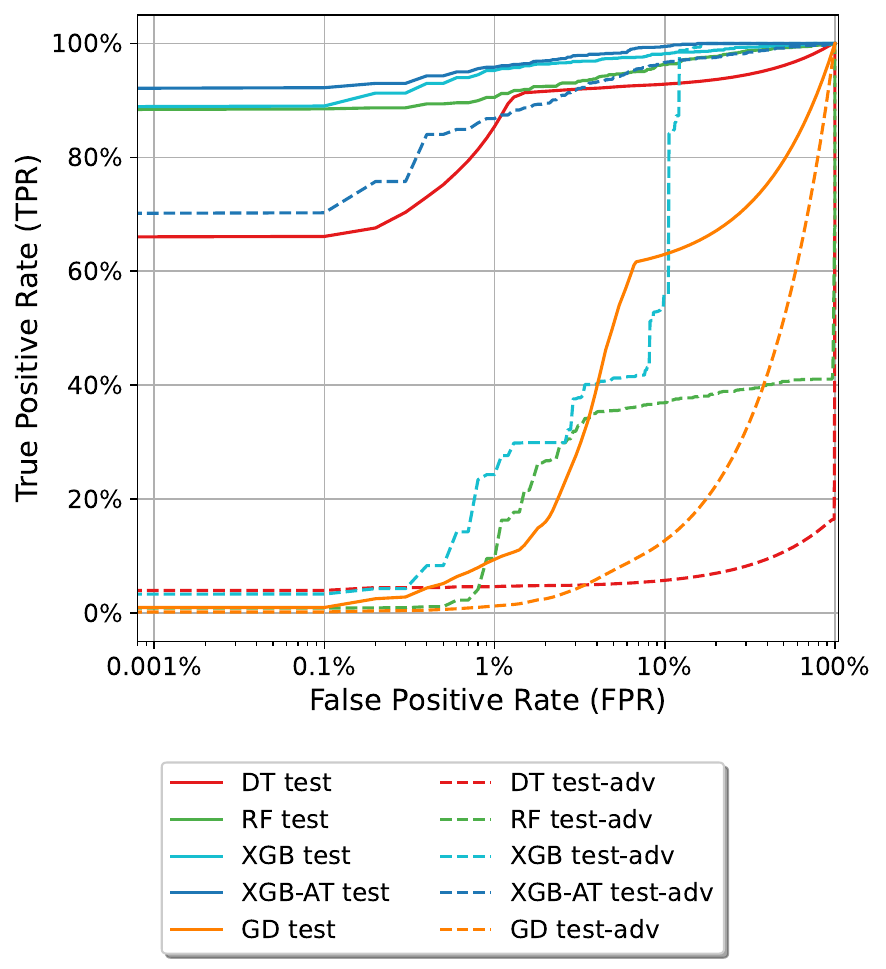}
    \caption{
    ROC curves of the detectors evaluated in this work on the baseline (\texttt{test}) and adversarial (\texttt{test-adv}) test sets,
    namely Decision Tree (\texttt{DT}), Random Forest (\texttt{RF}), XGBoost (\texttt{XGB}) and GuardDog (\texttt{GD}).
    The AT-based models are specified with \texttt{-AT}.
    }
    \label{fig:roc_detectors_all}
\end{figure}

In this section, we provide additional experimental results to complement those presented in Section~\ref{sec:results_baseline}.
Specifically, in Figure~\ref{fig:roc_detectors_all} we report the ROC curves of all the detectors evaluated on the baseline (\texttt{test}) and adversarial (\texttt{test-adv}) test sets created from \texttt{Malwarebench}.

\section{Feature Ablation Study}\label{sec:appendix-ablation-study}
In this section, we present the results of the feature ablation study conducted to evaluate the impact of the new features on the performance of the detector.

\newpage
\myparagraphlb{Baseline Evaluation.}
\autoref{tab:results_detectors_ablation} reports the recall (TPR) at 1\% FPR of the detectors based on the \texttt{original} (\ie, the original feature set proposed by Ladisa \etal~\cite{Ladisa2023ACSAC}) and \texttt{SoA} features,
evaluated on both the baseline (\texttt{test}) and adversarial (\texttt{test-adv}) test sets.
The results show that detectors based on the \texttt{SoA} features achieve a slightly higher recall on both test sets compared to those using the \texttt{original} features.
Specifically, the \texttt{SoA} features provide an average increase of 1.74\% in recall across all models evaluated on the \texttt{test} set.
This suggests that the \texttt{SoA} features enhance the detection capabilities of the models.

\myparagraphlb{Real-world Experiments.}
We then extended the ablation study to the \texttt{live1} real-world dataset.
\autoref{tab:results_live1_ablation} reports several performance metrics computed at 1\% FPR for the XGBoost detectors evaluated on \texttt{live1}.
The results highlight that, unlike in the baseline evaluation, the \texttt{SoA} features provide a significant improvement in the detection capabilities of the models on the real-world dataset,
with an average increase of 44.77\% in recall across both the baseline and AT-based models.
Furthermore, when combined with AT, the \texttt{SoA} features achieve the best performance:
the XGBoost model trained with AT and based on the \texttt{SoA} features outperforms all other configurations across all metrics and,
in particular, surpasses the baseline model with \texttt{original} features by 57.68\% in terms of recall.

\begin{table}[!t]
    \centering
    \begin{adjustbox}{max width=\columnwidth}
    \begin{tabular}{c c c c}
        \toprule
        \textbf{Detector} & \textbf{Features} & \multicolumn{2}{c}{\textbf{Dataset}} \\
        \cmidrule(lr){3-4}
        &  &  \texttt{test} & \texttt{test-adv} \\
        \toprule
        GuardDog             &  ---  &  6.63   &  1.00 \\
        \midrule
        \multirow{2}{*}{Decision Tree}  &  \texttt{original}   &  67.94  &  1.00 \\
                                        &  \texttt{SoA}    &  69.14  &  4.60 \\
        \midrule
        \multirow{2}{*}{Random Forest}  &  \texttt{original}   &  90.40  &  12.10 \\
                                        &  \texttt{SoA}    &  90.54  &  12.10 \\
        \midrule
        \multirow{2}{*}{XGBoost} &  \texttt{original}   &  93.00  &  14.00 \\
                                 &  \texttt{SoA}    &  95.27  &  24.30 \\
        \midrule
        \midrule
        \multirow{2}{*}{XGBoost AT}  &  \texttt{original}   &  93.20  &  85.87 \\
                                     &  \textbf{\texttt{SoA}}    &  \textbf{95.64}  &  \textbf{86.81} \\
        \bottomrule
    \end{tabular}
    \end{adjustbox}
    \vspace{+0.1cm}
    \caption{Recall (TPR) at 1\% FPR of the detectors evaluated on the (\texttt{test}) and (\texttt{test-adv}) sets.
    The last row reports the results of the XGBoost model trained with AT using 20\% of the adversarial packages.
    }
    \label{tab:results_detectors_ablation}
\end{table}

\begin{table}[!t]
    \centering
    \begin{adjustbox}{max width=\columnwidth}
    \begin{tabular}{c c c c c c c c c c c}
        \toprule
        \textbf{Model} &  \textbf{Features}  &  \textbf{FP} & \textbf{TP} & \textbf{FN} & \textbf{TN} & \textbf{Acc} & \textbf{Prec} & \textbf{Rec} & \textbf{F1} \\
        \toprule
        \multirow{2}{*}{base}  &   \texttt{original}  &   1218  &  156  &  198  &  120824  &  98.84  &  11.35  &  44.07  &  18.06 \\
                               &   \texttt{SoA}      &   1210  &  242  &  112  &  120832  &  98.92  &  16.67  &  68.36  &  26.80 \\
        \midrule
        \multirow{2}{*}{AT}  &   \texttt{original}  &  1215  &  183  &  171  &  120827  &  98.87  &  13.09  &  51.69  &  20.89 \\
                                   &   \textbf{\texttt{SoA}}      &  \textbf{1210}  &  \textbf{246}  &  \textbf{108}  &  \textbf{120832}  &  \textbf{98.92}  &  \textbf{16.90}  &  \textbf{69.49}  &  \textbf{27.18} \\
        \bottomrule
    \end{tabular}
    \end{adjustbox}
    \vspace{+0.1cm}
    \caption{Performance metrics at 1\% FPR of the XGBoost models based on the \texttt{original} and \texttt{SoA} features, evaluated on the \texttt{live1} dataset.
    \texttt{base} represents the baseline models, while \texttt{AT} represents the model trained with AT using 20\% of adversarial samples.
    }
    \label{tab:results_live1_ablation}
\end{table}

\newpage
\bibliographystyle{IEEEtran}
\bibliography{bibliography.bib}

\end{document}